\documentclass[fleqn,usenatbib]{mnras}

\usepackage{newtxtext,newtxmath}
\usepackage[T1]{fontenc}
\usepackage{float}
\DeclareRobustCommand{\VAN}[3]{#2}
\let\VANthebibliography\thebibliography
\def\thebibliography{\DeclareRobustCommand{\VAN}[3]{##3}\VANthebibliography}

\usepackage{graphicx, subcaption}	
\usepackage[export]{adjustbox}
\usepackage{amsmath}	% Advanced maths commands
\usepackage{xcolor}
\usepackage{tabularx}
\usepackage{threeparttable, tablefootnote}
\usepackage{longtable}
\usepackage{lscape}
\usepackage{subcaption}
\usepackage{multirow}

\newcommand{\hii}{\mbox{\ion{H}{II~}}} 
\title[Brown dwarf Analysis in IC 1396]{Search for Brown Dwarfs in IC 1396 with Subaru HSC: Interpreting the Impact of Environmental Factors on Sub-stellar Population}

\author[Gupta et al.]{Saumya Gupta $^{1}$\thanks{kcsaumya.gupta@gmail.com}, 
Jessy Jose$^{1}$\thanks{jessyvjose1@gmail.com}, 
Swagat R. Das$^{2}$, 
Zhen Guo$^{3,4,5,6}$, 
Belinda Damian$^7$,
\newauthor
Prem Prakash$^8$,
Manash R Samal$^9$,
\\
$^{1}$ Indian Institute of Science Education and Research (IISER) Tirupati, Rami Reddy Nagar, Karakambadi Road, Mangalam (P.O.), Tirupati 517 507, India\\
$^{2}$ Departamento de Astronom{\'i}a, Universidad de Chile, Las Condes, 7591245 Santiago, Chile\\
$^{3}$Instituto de F{\'i}sica y Astronom{\'i}a, Universidad de Valpara{\'i}so, ave. Gran Breta{\~n}a, 1111, Casilla 5030, Valpara{\'i}so, Chile\\
$^{4}$N\'ucleo Milenio de Formaci\'on Planetaria (NPF), ave. Gran Breta{\~n}a, 1111, Casilla 5030, Valpara{\'i}so, Chile\\
$^{5}$Centre for Astrophysics Research, University of Hertfordshire, Hatfield AL10 9AB, UK\\
$^{6}$Departamento de F{\'i}sica, Universidad Tecnic{\'a} Federico Santa Mar{\'i}a, Avenida Espa{\~n}a 1680, Valpara{\'i}so, Chile\\
$^{7}$ Christ ({\it Deemed to be University}), Bangalore, India\\
$^{8}$ Indian Institute of Technology Hyderabad, Kandi, Sangareddy, Telangana, India\\
$^{9}$ Physical Research Laboratory (PRL), Navrangpura, Ahmedabad 380 009, Gujarat, India\\
}

\date{Accepted XXX. Received YYY; in original form ZZZ}

\pubyear{2024}

\begin{document}
\label{firstpage}
\pagerange{\pageref{firstpage}--\pageref{lastpage}}
\maketitle

\begin{abstract}
Young stellar clusters are predominantly the hub of star formation and hence, ideal to perform comprehensive studies over the least explored sub-stellar regime. Various unanswered questions like the mass distribution in brown dwarf regime and the effect of diverse cluster environment on brown dwarf formation efficiency still plague the scientific community. The nearby young cluster, IC 1396 with its feedback-driven environment, is ideal to conduct such study. In this paper we adopt a multi-wavelength approach, using deep Subaru HSC, Gaia DR3, Pan-STARRS, UKIDSS/2MASS photometry and machine learning techniques to identify the cluster members complete down to $\sim$ 0.03 M$_{\odot}$ in the central 22$^{\prime}$ area of IC 1396. We identify 458 cluster members including 62 brown dwarfs which are used to determine mass distribution in the region. We obtain a star-to-brown dwarf ratio of $\sim$ 6 for a stellar mass range 0.03 -- 1 M$_{\odot}$ in the studied cluster. The brown dwarf fraction is observed to increase across the cluster as radial distance from the central OB-stars increases. This study also compiles 15 young stellar clusters to check the variation of star-to-brown dwarf ratio relative to stellar density and UV flux ranging within 4-2500 stars pc$^{-2}$ and 0.7-7.3 G$_{0}$ respectively. The brown dwarf fraction is observed to increase with stellar density but the results about the influence of incident UV flux are inconclusive within this range. This is the deepest study of IC 1396 as of yet and it will pave the way to understand various aspects of brown dwarfs using spectroscopic observations in future. %This is crucial to enhance general understanding about the role of external factors on brown dwarf population. 
\end{abstract}


%%%%%%%%%%%%%%%%%%%%%%%%%%%%%%%%%%%%%%%%%%%%%%%%%%

%%%%%%%%%%%%%%%%% BODY OF PAPER %%%%%%%%%%%%%%%%%%

\section{Introduction}
\label{sec: intro}
Brown dwarfs, defined as self-gravitating sub-stellar objects with mass ranging between 0.013 -- 0.08 M$_\odot$, have cool temperatures insufficient to initiate Hydrogen burning in their cores. Several mechanisms proposed for brown dwarf formation include turbulent fragmentation, disk fragmentation, dynamical ejection and photo-erosion (\citealt{2004A&A...427..299W, 2008MNRAS.389.1556B, 2009MNRAS.392..413S, 2018arXiv181106833W}). These sub-stellar sources have been the focus of several past and recent studies in different young star-forming regions such as ONC, NGC 2244, Lupus, $\rho$ Oph, W3, Taurus, Perseus, Sigma Ori, Serpens, Upper Sco and IC 348 (\citealt{2012ApJ...744....6S, 2017ApJ...842...65Z, 2017MNRAS.471.3699M, Luhman_2018, 2019AJ....158...54E, 2019ApJ...881...79M, 2020ApJ...892..122J, 2021AJ....161..138H, 2023A&A...677A..26A, 2023JApA...44...77D}). In spite of such plethora of studies, the exact dominant brown dwarf formation mechanism and the factors which determine the brown dwarf population in a region still remain vague. Many brown dwarf formation mechanisms suggest that the pre-mature ejection of a self-gravitating object results in the formation of very low-mass objects as they are cut-off from the accretion material supply (\citealt{2004A&A...427..299W, 2009MNRAS.392..413S, 2014prpl.conf..619C, 2018arXiv181106833W}). It is hence, expected that the external factors like turbulence, high energy UV radiation and stellar density may introduce more dynamical perturbations leading to an early ejection of the collapsing cores and enhance the census of brown dwarfs formed in a region. Another interesting yet ambiguous subject to investigate here is the mass distribution in sub-stellar regime. Deep studies of Galactic stellar clusters reaching down to the low mass and brown dwarf realm are helpful to constrain the stellar evolutionary models. It is thus, pivotal to obtain robust statistical samples of brown dwarfs in various young stellar clusters ($<$ 10 Myrs) with diverse environments to shed more light on the ambiguous issues related to the sub-stellar realm.

The young stellar clusters ($<$ 10 Myrs) are not much dynamically evolved (\citealt{2010ARA&A..48..431P, 2014prpl.conf..291L, 2016ApJ...817....4F}), hence their current mass distribution is closest to the initial mass function (IMF) of the cluster (\citealt{2002MNRAS.336.1188K, 2015arXiv151101118L}). Such young clusters provide a robust sample of stars spanning a wide range of mass and hence, are the ideal test-beds to perform deep comprehensive studies of the stellar and sub-stellar population in the region. However, due to obvious observational constraints and high intrinsic reddening from the cluster, it is difficult to observe the faint low-mass and sub-stellar population ($<$ 0.5 M$_\odot$) in distant ($>$ 1 kpc) young Galactic star forming regions. On the contrary, the nearby young clusters affected by minimal extinction like ONC, $\rho$ Ophuichi, IC 348 and NGC 1333 (A$_V$ $\sim$ 2 mag) provide excellent statistical sample to carry out deeper and wider studies in diverse environments. In addition, although spectroscopy is the widely adopted solution of obtaining the stellar parameters in nearby clusters, it is highly time intensive and restricted to a limited number of sources (\citealt{2012A&A...539A.151A, 2019BAAS...51c.547B, 2021MNRAS.507.4074P, 2023ApJ...951..139D}). Hence, multi-wavelength photometric studies of such star forming regions with facilities like Pan-STARRS, Gaia, Subaru Hyper Suprime-Cam (HSC), GTC, HST, CFHT/WIRCam, and UKIRT/WFCAM are useful to detect the very low-mass end of stellar population (\citealt{2012ApJ...744..134M, 2020PASP..132j4401A, 2020ApJ...896...79R, 2023ApJ...948....7D, 2023ApJ...951..139D}).\\ 

It is also crucial to understand the effect of diverse environmental conditions on the population of brown dwarfs in a cluster. Brown dwarf fraction is a useful tool to estimate the statistics of brown dwarfs formed in a given star forming event. Any increase or decrease in the brown dwarf formation efficiency across diverse cluster environments due to external factors like UV flux and stellar density, would imply an inverse effect on the star-to-brown dwarf ratio (\citealt{2004A&A...427..299W, 2008MNRAS.389.1556B, 2009MNRAS.392..413S, 2017A&A...608A.107V}). The feedback driven cluster environment may promote the formation of brown dwarfs in a region by depleting the supply of accreting material from the fragmented low-mass cores. Recent studies have also explored the impact from the presence of OB stars on the formation of brown dwarfs (\citealt{2019ApJ...881...79M, 2023A&A...677A..26A}) for the regions covered under the Sub-stellar Objects in Nearby Young Clusters (SONYC) survey. In the present study, we identify and characterize the brown dwarfs in a young star forming region, IC 1396 (see details below), and then compare the star-to-brown dwarf ratio with various star-forming regions. We also compile fifteen young stellar clusters to investigate the behaviour of star-to-brown dwarf ratio with varying stellar density and incident FUV flux.

%This paper is divided into the following sections: The Section Fig \ref{sec:data} interprets the Subaru Hyper Suprime-Cam observations, data reduction and catalog generation using HSC pipeline. Section Fig \ref{sec:quality} presents the data quality in terms of photometry, astrometry, completeness of the HSC data along with comparison relative to already available optical photometry. In  Section Fig \ref{sec:analysis} we present the data analysis and results obtained, aided with colour-magnitude diagrams, age analysis and disk fraction analysis. We then discuss and interpret the results obtained with this data so far in Section Fig \ref{sec:discuss} and encapsulate the entire work along with our future plans, finally in Section Fig \ref{sec: sumup}.\\

\subsection{IC 1396}
\label{sec: IC 1396}

\begin{figure}
	\centering
	\includegraphics[scale = 0.32]{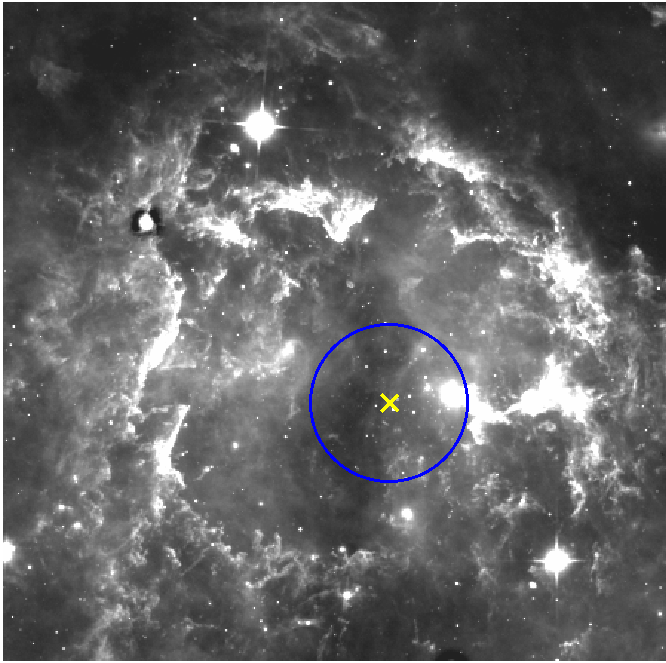}
	%\includegraphics[scale = 0.38]{r2-Y_vs_r2.eps}%\scriptsize
	%\includegraphics[scale = 0.4]{z-K_vs_z.eps}%\scriptsize
	%\begin{small}  
	\scriptsize
	\linespread{0.8}
	\caption {WISE 22 $\mu$m image of IC 1396 with the blue circle marking the 22$^\prime$ radius region studied in this paper. The yellow cross marks the massive star HD 206267. 
	 }
%{\it Top left:}
% }
%\end{small}
\label{fig:IC 1396}
\end{figure}

The young star forming \hii region IC 1396 (see Figure \ref{fig:IC 1396}), almost 3$^\circ$ in diameter, is associated with the Cepheus OB2 association (\citealt{1998ApJ...507..241P, 2007ApJ...654..316G}). At a distance of 900 -- 910 parsecs %900$^{+100}_{-50}$ parsecs 
(\citealt{2002AJ....124.1585C,2006ApJ...638..897S, 2019A&A...622A.118S, 2023ApJ...948....7D}), this region is ionised by the central binary system HD 206267 (\citealt{2023A&A...669A..22P} and references therein) and is estimated to have an age ranging between 2--5 Myrs (\citealt{1998ApJ...507..241P, 2005AJ....130..188S, 2012MNRAS.426.2917G, 2013A&A...559A...3S}). The cluster hosts a prominent cavity at the center cleared off the dust and gas as a result of ionization from the central massive star. The region is also associated with several Bright-rimmed clouds (BRCs), globules and elephant-trunk structures. The relatively close proximity, minimal uniform extinction (A$_V$ $\sim$ 1 -- 1.5 mag (\citealt{2002AJ....124.1585C, 2005AJ....130..188S, 2012AJ....143...61N, 2013A&A...559A...3S, 2019A&A...622A.118S})) and ongoing star formation affected by the feedback from massive stars in the region (\citealt{2004AJ....128..805S, 2006ApJ...638..897S, 2012MNRAS.421.3206S}), make it an ideal target to conduct deep studies regarding low-mass star formation and the factors affecting it.

Several studies have performed the membership classification of the region using techniques like X-ray observations (\citealt{2009AJ....138....7M, 2012MNRAS.426.2917G}), spectroscopic surveys (\citealt{2002AJ....124.1585C, 2006ApJ...638..897S, 2012AJ....143...61N, 2013A&A...559A...3S, 2019A&A...622A.118S}), H$\alpha$ photometry (\citealt{2011MNRAS.415..103B}) and infrared disk searches (\citealt{2006ApJ...638..897S, 2009ApJ...702.1507M, 2021AJ....162..279S}). The most recent studies of IC 1396 by \citet{2023A&A...669A..22P} and \citet{2023ApJ...948....7D} add extra $\sim$ 500 and 244 new members respectively, to this list using Gaia astrometry. In total, all these studies identify 2455 sources as members in the entire 3$^{\circ}$ diameter region. However, none of these studies are complete for stellar masses $<$ 0.5 M$_\odot$ and hence, are inadequate to explore the sub-stellar population in the region. This calls for deeper observations using a wide field-of-view camera on a large telescope which can simultaneously reach down to faint low mass limits while covering a wider area.

This study aims to perform a brown dwarf population analysis within the 22$^\prime$ region, centered at {\it R.A}: 324.74$^{\circ}$ ; {\it Dec}: 57.49$^{\circ}$ around the massive star HD 206267. The area is specifically chosen for study as more than 60$\%$ of the previously identified cluster members as well as disk sources are located within this area. The complex is assumed to have multiple generations of stellar population in BRCs (\citealt{2006ApJ...638..897S,  2014A&A...562A.131S, 2023A&A...669A..22P, 2023ApJ...948....7D}) and we focus on the central portion of the cluster.\\ %Also, \citet{2005A&A...432..575F} observe an inverse correlation between stellar density and distance from the massive O star HD 206267 in IC 1396. These authors find that most of the globules with high stellar density lie within 0.17 -- 0.2 pc from HD 206267. The considered area is hence, expected to be rich in sub-stellar population as well. In this work, we identify the brown dwarf members, analyse their properties and determine the mass distribution down to the brown dwarf regime in the 22$^\prime$ radius region, . We plan to extend this analysis to a larger area in future.\\

This work is a premier study to investigate the brown dwarf population in IC 1396. It presents a compiled comparative analysis of fifteen star forming regions, which will aid the current colloquy about the role played by external factors on brown dwarf formation efficiency. This paper is organized as follows: Section \ref{sec:data} presents the multi-wavelength data sets used for this study. Section \ref{sec:method} discusses the various machine learning techniques, spectral energy distribution (SED) fitting and pre-main sequence locus based identification to obtain a reliable set of members. Consequently, we identify the sub-stellar sources and determine the mass distribution in Section \ref{sec:analysis}. We perform a comparative study of the brown dwarf fraction across fifteen star forming regions and discuss its variation relative to stellar density and incident FUV flux in Section \ref{sec:discuss}. A summary of our work along with the future plans are presented in Section \ref{sec: sumup}. Finally, we present the identified brown dwarfs in Section \ref{sec: bd_table}.

\section{\textbf{Data Sets Used}}
\label{sec:data}
The multi-wavelength broadband photometry is a useful approach to detect a substantial sample of sub-stellar objects in nearby young star forming regions (\citealt{2013A&A...549A.123A, Esplin_2017, Luhman_2018}). However, the multi-wavelength photometry approach also requires a reliable member sample which can be used as a template for other sources. The Gaia astrometry including proper motion and parallax values for individual sources is ideal to generate such a model member sample. This can be further applied to obtain a deeper set of members using machine learning (ML) techniques.  %Although spectroscopy is more reliable and widely used method to estimate the stellar parameters, it is restricted to a limited sample of sources. The deep multi-wavelength photometry is particularly helpful and sufficient to obtain a large sample of low mass sources especially in nearby star-forming regions like IC 1396 which is affected by minimal amount of uniform extinction (A$_{V}$ $\sim$ 1--1.5 mag) and hence, do not require confirmation by follow-up spectroscopy. 
We use optical (Subaru HSC, Pan-STARRS DR1 and Gaia DR3) in conjunction with near-IR (NIR) (UKIDSS DR11, 2MASS) data to perform a deep study of the brown dwarf realm of the target region.\\

\subsection{Subaru HSC Data}
%\label{hsc_data}
We obtained Subaru HSC observations for two pointings (central {\it RA:} 324.27$^{\circ}$; {\it Dec:} 57.91$^{\circ}$ and {\it RA}: 326.37$^{\circ}$; {\it Dec:} 57.72$^{\circ}$, respectively), each covering an area of 1.5$^\circ$ diameter. The HSC observations were taken on 18th August'2017 (PI: J. Jose; Program ID: S17B0108N), using EAO\footnote{East Asian Observatory} time in three broad-band optical filters, namely, r$_{2}$, i$_{2}$ and Y (\citealt{2018PASJ...70...66K}). The data reduction is conducted following the steps designed in \citet{2021MNRAS.508.3388G}. The raw data is reduced by HSC pipeline version 6.7 which performs (1) Single-visit Processing which involves detrending of the raw data (2) Joint Calibration which conducts an internal calibration across different visits\footnote{observing shots} (3) Coaddition of various visits to form a single deep image and finally, (4) Coadd Processing/ Multiband Analysis which detects individual sources using source extraction and performs photometry on coadded images in the individual filters. For more details on the HSC data reduction please refer \citet{2018PASJ...70S...8A, 2018PASJ...70S...5B, 2021MNRAS.508.3388G}.\\
We obtain a total of 2,401,026 sources detected in either of the three filters (r$_{2}$, i$_{2}$ or Y), after data reduction for the entire observed region. However, we will only use the data restricted within the radius of 22$^\prime$ around the ionizing source HD 206267 for our analysis due to the aforementioned reasons (see Section \ref{sec: IC 1396}). The photometric results and analysis of a wider area of the complex will be presented in a forthcoming paper (Das et al., in preparation). Approximately 254,022 HSC sources are located within the area of interest with photometry in either of the three bands. In order to select sources with good photometry as well as avoid losing genuine sources, we consider only those which have photometric error $\leq$ 0.1 mag in individual HSC filters (\citealt{2021MNRAS.508.3388G}). The data is observed to saturate at r$_{2}$ $\sim$ 15.0 mag corresponding to a source of $\sim$ 1 M$_\odot$, and is 90$\%$ complete down to 24.5 mag which corresponds to a 0.03 - 0.035 M$_\odot$ source, at a distance of 900 parsecs, age $\sim$ 2 Myrs and A$_{V}$ $\sim$ 1 mag (\citealt{2015A&A...577A..42B}). We obtain the 90$\%$ completeness limit here using the turnover point of source count approach (for details see \citet{2017ApJ...836...98J, 2021MNRAS.504.2557D, 2021MNRAS.508.3388G}). With the help of such deep optical data, we are able to explore well into the sub-stellar realm as discussed in Section \ref{sec:analysis}.\\

\subsection{Pan-STARRS DR1 Data}
\label{sec:ps_data}
In addition to HSC, we use Pan-STARRS DR1 data\footnote{downloaded from https://vizier.u-strasbg.fr/viz-bin/VizieR} (\citealt{2016arXiv161205560C}) in r, i and Y filters to generate the optical catalog used for our analysis. We consider only those sources for which number of stack detections is $>$ 2, magnitude in individual bands $\geq$ 12 mag to avoid source saturation and photometric error in individual filters is $\leq$ 0.1 mag (\citealt{2016arXiv161205560C}). We use transformation equations given in \citet{2021MNRAS.508.3388G} to convert the photometry from Pan-STARRS filter system to HSC system. Pan-STARRS photometry is mainly used here to add sources, especially at the bright end, which are missed in HSC photometry. %We use the cross-matching radius of 0.1$^{\prime\prime}$ to search for counterparts between Pan-STARRS and HSC. 
In total, we have a combined optical catalog of 258,602 sources which will be further used to identify members in the area of study. We will refer this combined data set as the deep optical catalog henceforth.\\

\subsection{Gaia DR3 Data}
\label{sec:gaia data}
We obtained the Gaia DR3 data consisting of 48,185 sources from Gaia Science archive (\citealt{2022arXiv220800211G}) in the 22$^\prime$ radius of IC 1396 and use it to obtain a reliable training set for the identification of members using machine learning techniques as elaborated in further sections. In order to select genuine Gaia sources, we give certain constraints such as parallax $>$ 0, re-normalised unit weight error (RUWE) $<$ 1.4, detection in all Gaia filters (\citealt{2021A&A...649A...2L, 2022arXiv221011930P, 2022A&A...664A.175P}) while retrieving the data using Astronomical Data Query Language (ADQL) interface. Since, Gaia data is complete between 19 $\leq$ G $\leq$ 20 mag (\citealt{2021A&A...649A...1G, 2022arXiv220605796M}), we restrict the sources down to G $\leq$ 19.5 mag (average of range 19--20). We thus, obtain 19,432 Gaia sources within the 22$^\prime$ radius of IC 1396.\\ 

\subsection{UKIDSS GPS DR11 / 2MASS Data}
The NIR photometry when used along with deep optical data, gives a broad wavelength range particularly useful for performing SED fitting of the cluster members and obtaining the relevant stellar parameters (\citealt{2019ApJ...881...79M, 2021MNRAS.504.2557D, 2021A&A...650A..48K}). We use the deep archived UKIDSS GPS DR11\footnote{http://wsa.roe.ac.uk/dr11plus\_release.html} (\citealt{2008MNRAS.391..136L}) in J, H and K bands (\citealt{2006MNRAS.367..454H, 2007MNRAS.379.1599L}) for this purpose. The UKIDSS GPS DR11 data is retrieved from WFCAM Science archive after applying constraints like J $\geq$ 13 mag and photometric errors less than 0.1 mag to only obtain sources with good photometry. In order to include brighter sources, we make use of the 2MASS NIR data (\citealt{2003tmc..book.....C}) downloaded from Vizier in JHK bands with J $<$ 13. Also, we select good 2MASS sources based on the quality flags such as ph$\_$qual $\neq$ F,E,U, rd$\_$flg $\neq$ 6 and cc$\_$flg $\neq$ p,d,s,b (\citealt{2003tmc..book.....C, 2021A&A...650A.157G}). The concatenated list of sources from 2MASS (J $<$ 13 mag) and UKIDSS (J $\geq$ 13 mag) forms the deep NIR catalog of the central 22$^{\prime}$ radius region of IC 1396, comprising of 87,614 sources to be used further for our analysis.\\

%\footnotetext[12]{Error in magnitude in individual HSC filters}

\section{\textbf{Membership analysis}}
\label{sec:method}
The identification of cluster members based on the stellar proper motion and parallax information is one of the most reliable and widely used methods (\citealt{2019A&A...625A.115O, 2019A&A...631A..57M, 2021A&A...650A..48K, 2022A&A...664A.175P, 2023A&A...669A..22P, 2023ApJ...948....7D}). Thanks to the consecutive Gaia data releases, we now have complete astrometric solutions (proper motion and parallax) available for 1.46 billion stars (\citealt{2022arXiv220800211G}). However, the primary drawback here is that it is only complete down to G $\sim$ 19 - 20 mag (\citealt{2021A&A...649A...1G, 2022A&A...664A.175P, 2022arXiv220605796M}) and hence, quite shallow to identify the faint low-mass cluster members. %The deep exhaustive studies of complete sample of members, especially towards the low mass and sub-stellar end, in young star forming regions can help in constraining the mass distribution, dynamical evolution and hence star formation history of the region. 
Machine learning is an  extensively used tool which comes to our rescue at this juncture with its unsupervised and supervised techniques like Gaussian Mixture Model (GMM) and Random Forest classifier (RF). Such techniques can accurately classify cluster members and young stellar objects from non-members in young star forming regions based on the input features (\citealt{2014A&A...563A..45S, 2018ApJ...869....9G, 2019PASP..131d4101G, 2019A&A...625A.115O, 2020AJ....159..200M, 2023A&A...669A..22P, 2023ApJ...948....7D}).\\ 

In this study, we use Gaia archived data to form a set of training data by clustering them into members and non-members using the GMM technique. Reliable members and non-members are then utilized in training the RF classifier to identify the probable members among the input deep optical data set. This is followed by SED analysis of the probable members to select confirmed members and deduce their stellar parameters. We briefly describe the procedure adopted to identify members in the area of interest, down to very faint low-mass and sub-stellar sources ($\sim$ 0.03 M$_\odot$). Please refer \citet{2023ApJ...948....7D} for more details about the procedure followed.

\subsection{GMM}
\label{sec:gmm}
GMM is an unsupervised machine learning technique which categorizes a data set into different clusters based on the probability of each data point associated with a particular group. The technique makes a reasonable assumption that the data set is a collection of normally distributed groups (\citealt{2014A&A...563A..45S, 2018Ap&SS.363..232G, 2018ApJ...869....9G, 2019A&A...624A.126C, 2020A&A...643A.148G}). We initiate our cluster member selection process by generating an efficient training set using Gaia DR3 data for the 22$^\prime$ radius region. However, one of the major limitations of this method is failure to identify clusters due to overcrowding of field stars (\citealt{2018Ap&SS.363..232G, 2018ApJ...869....9G, 2023ApJ...948....7D}). In order to avoid this, we impose few more constraints on astrometric parameters such as proper motion values in the range --20 $\leq$ $\mu_{\alpha}$cos$\delta$ $\leq$ 20 mas/yr , --20 $\leq$ $\mu_{\delta}$ $\leq$ 20 mas/yr (\citealt{2013AN....334..673E}) and distance range within 500 parsecs -- 1500 parsecs (\citealt{1999AJ....117..354D, 2023A&A...669A..22P}) while using the Gaia data here. We use the distance estimates from \citet{2021AJ....161..147B}. The mentioned constraints on the distance and proper motions are relaxed enough so as not to lose any probable members as well as ensure reduction in the foreground and background contamination in accordance with the values estimated by the previous studies. This aids in generating a robust training data set.\\

\begin{figure*}
	%\centering
	\includegraphics[scale=0.2]{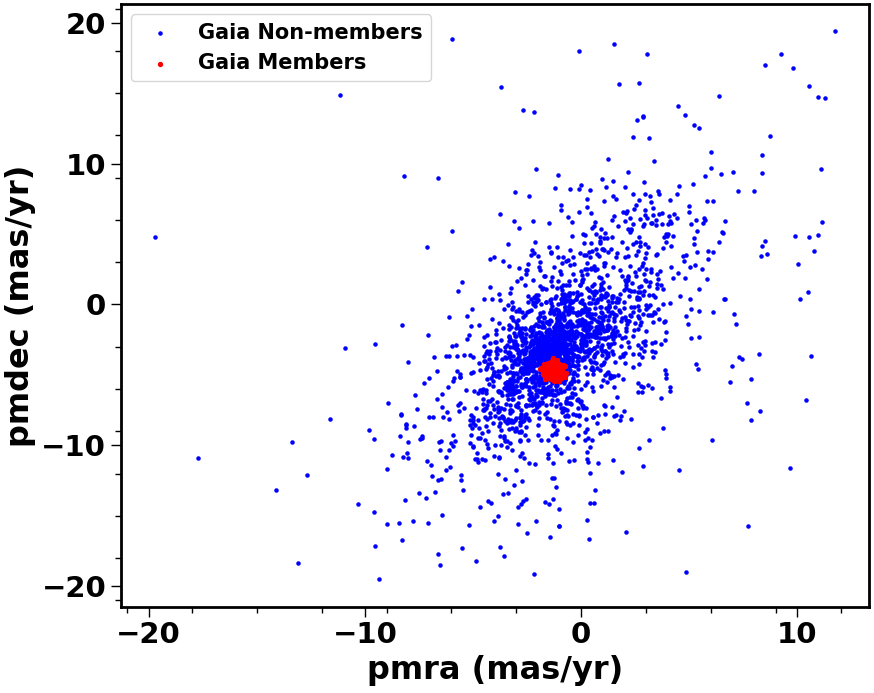}
	\includegraphics[scale=0.205]{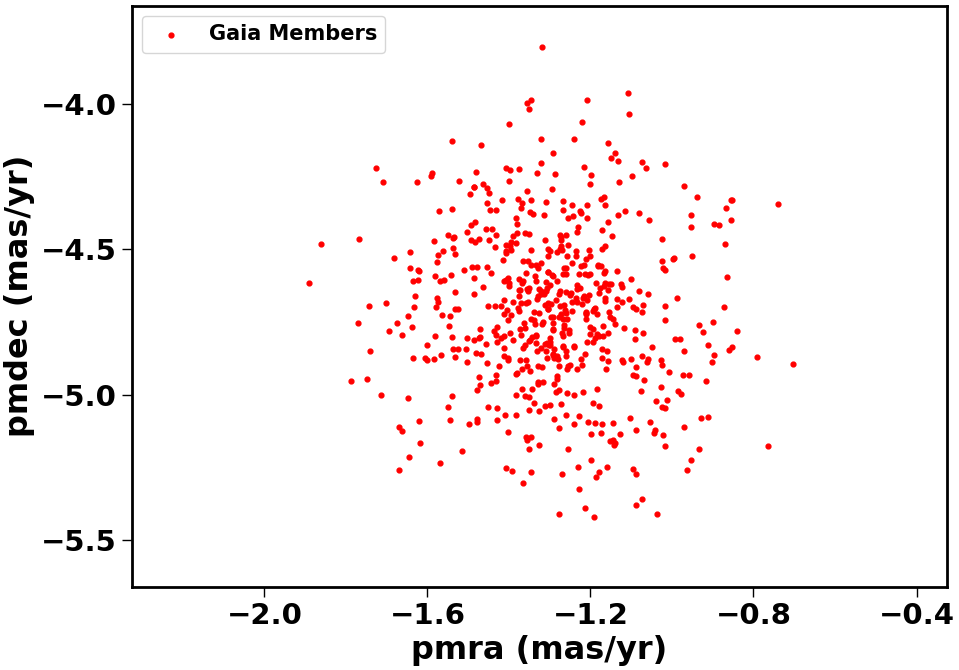}
	%\includegraphics[scale=0.18]{Del_Dec_spatial.png}
	%\includegraphics[scale = 0.2]{cyg_centre1.eps}
	%\includegraphics[scale = 0.38]{r2-Y_vs_r2.eps}%\scriptsize
	%\includegraphics[scale = 0.4]{z-K_vs_z.eps}%\scriptsize
	%\begin{small}  
	%\scriptsize
	%\linespread{0.8}
	\caption{({\it Left}) Proper motion RA vs proper motion Dec plot obtained after performing GMM. ({\it Right) Zoomed version of proper motion RA vs proper motion Dec plot focused on GMM identified members.}
	 }%{\it Top left:}
% }
%\end{small}
\label{fig:pmra}
\end{figure*}

\begin{figure*}
	%\centering
	\includegraphics[scale=0.24]{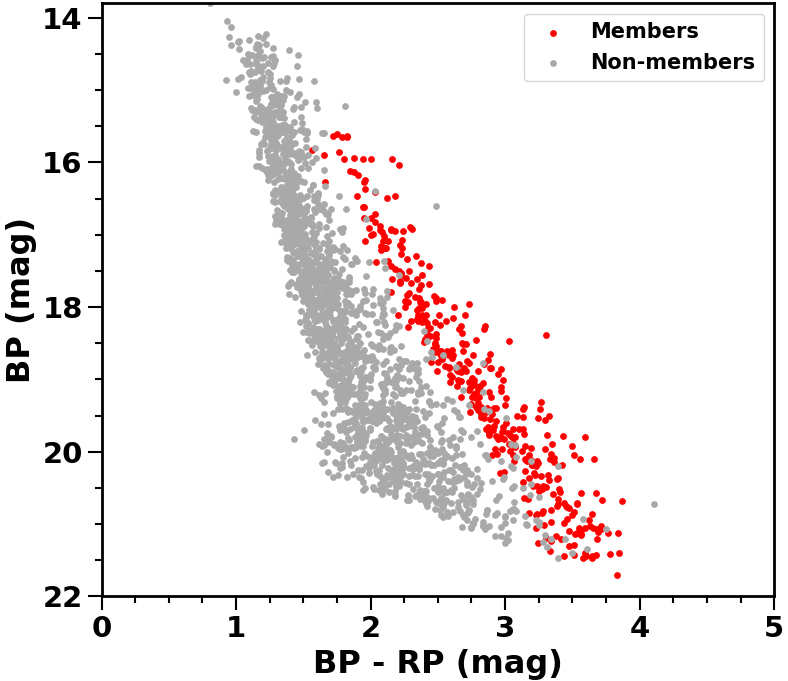}
	\includegraphics[scale=0.235]{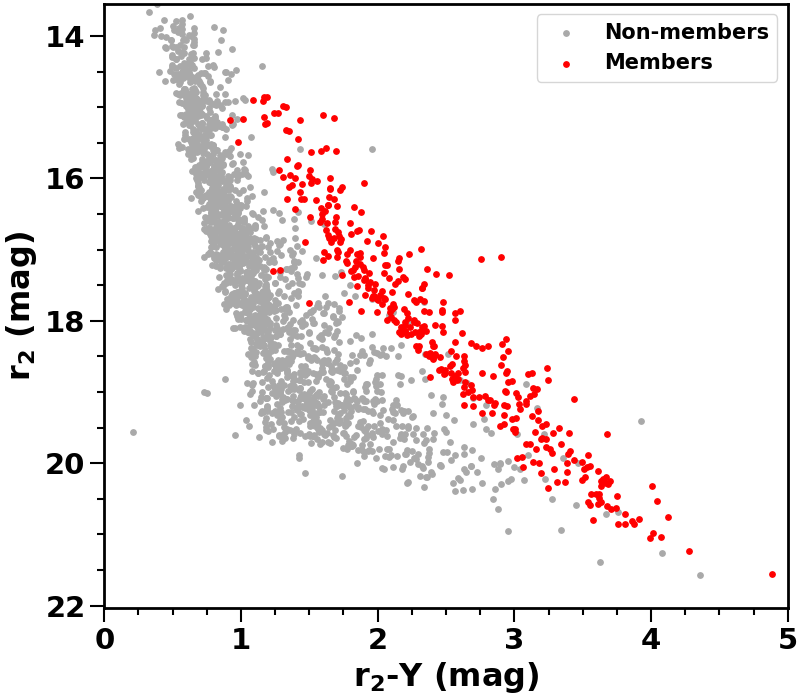}
	%\includegraphics[scale=0.18]{Del_RA_spatial.png}
	%\includegraphics[scale=0.18]{Del_Dec_spatial.png}
	%\includegraphics[scale = 0.2]{cyg_centre1.eps}
	%\includegraphics[scale = 0.38]{r2-Y_vs_r2.eps}%\scriptsize
	%\includegraphics[scale = 0.4]{z-K_vs_z.eps}%\scriptsize
	%\begin{small}  
	%\scriptsize
	%\linespread{0.8}
	\caption{ ({\it Left}) BP - RP vs BP ({\it Right}) r$_{2}$ - Y vs r$_{2}$ CMD for members and non-members obtained after performing GMM.
	 }%{\it Top left:}
% }
%\end{small}
\label{fig:train_cmd}
\end{figure*}

In total, we obtain 2771 Gaia sources after applying the above constraints. Among these stars, GMM identifies two groups, namely members and non-members and assigns each star the probability of being associated to each group. This assignment is based on astrometric parameters, i.e, positions, proper motions and parallax information of the star. The sources identified as members and with probability $\geq$ 0.8 (\citealt{2019PASP..131d4101G}) are selected for further analysis. The sources identified as non-members by GMM based on their astrometric information, are used as non-member training set further in RF classifier. %In order to select only reliable sources for the training set, we consider only those member sources for which the probability score is $\geq$ 0.8, which constitutes $\sim$ 93$\%$ ($\sim$ 613 sources) of the labelled members. 
Since IC 1396 is a young cluster with age $<$ 10 Myrs (\citealt{2005AJ....130..188S, 2023ApJ...948....7D}), we further refine the Gaia member selection by considering only those sources which are located to the right of the 10 Myr isochrone, corrected for a distance $\sim$ 900 parsecs and A$_{V}$ $\sim$ 1 mag (\citealt{2015A&A...577A..42B}) in all the colour-magnitude combinations ((G--RP) vs G, (BP--RP) vs BP, (BP--G) vs BP). We present the pmra vs pmdec plots, BP - RP vs BP and r$_{2}$ - Y vs r$_{2}$ CMD in Fig \ref{fig:pmra} and Fig \ref{fig:train_cmd} for the identified Gaia-based members. We eventually secure 411 Gaia members which form an ideal training set for RF classifier.\\ %We present pmra vs pmdec plot in Fig \ref{fig:pmra} and (BP--RP) vs BP colour-magnitude diagram (CMD) overplotted with Gaia members in Fig. The CMD clearly shows that these members are well segregated from the non-members, thus forming an ideal training data set.\\

\subsection{Random Forest Classifier}
\label{sec:rf}
RF classifier is a widely used supervised learning technique (\citealt{2020A&A...643A.148G, 2022A&A...668A..19M, 2023ApJ...948....7D}) which performs classification of a new data set based on the training data by using the majority voting approach. The algorithm builds individual decision trees which generate output for each of the different sample subsets of the data. The eventual output of the classification is considered based on the results of the majority of the subsets (\citealt{2011JMLR...12.2825P}). Due to this approach, RF classifier is immune to dimensionality of the feature\footnote{input attributes useful in learning distinction between different classes} space. We identify probable members among the deep optical data by executing RF classifier in three phases. We use the above identified Gaia members and non-members with HSC counterparts (see Section \ref{sec:gmm}) as the training data set. In the first phase, RF classifier is applied to the 34,859 HSC sources with Gaia DR3 counterparts devoid of any constraints applied to it. In this phase, we are able to use Gaia parallax and proper motions along with the optical colours ((r$_{2}$--i$_{2}$) vs r$_{2}$ ; (r$_{2}$--Y) vs r$_{2}$) and magnitude ratios (Y/r$_{2}$ ; i$_{2}$/r$_{2}$) of stars as input training features. %\textbf{The i$_{2}$ and Y combination of color and ratio are not used for training purpose. This is because this filter combination biases the RF model by predicting more of the majority class, which in any data set, is the non-member sample. A biased RF model assigns higher importance ($>$ 0.5) to such feature. Hence, using (i$_{2}$ - Y) and (Y/i$_{2}$) as input features diminishes the quality of segregation between members and non-members by the RF classifier.}
The RF model, thus trained on the basis of training features mentioned above identifies 792 probable members. %Since i$_{2}$ and Y filters have the maximum number of detections (them being the longest wavelengths among HSC optical filters), it consequently predicts a higher proportion of non-members is correspondingly higher 

In the second phase, we use this obtained member and non-member set as the training sample for the remaining HSC sources without Gaia counterparts and hence, devoid of any parallax and proper motion information of the sources. The RF classifier in this phase uses the same optical colours and magnitude ratios as mentioned above for the training purpose. Using magnitude ratios along with optical HSC colours as input features aids in improving the quality of segregation between the members and non-members, as evident from the receiver operating characteristic (ROC) curves (shown in Appendix \ref{sec:rf_quality}). The hyper-parameters used for RF classifier are also listed in Appendix \ref{sec:rf_quality}. For each run of RF classifier, we evaluate the model using train-test-split with test size equal to 40$\%$ of the data set and random state equal to 50. We train our algorithm using 5-fold cross-validation with an average cross validation (CV) score of 0.985 and using ROC as the measure of performance. %The estimated accuracy of the probable candidate members in each run of the RF classifier is $\sim$ 98$\%$ - 99$\%$, as evident from the confusion matrix shown in Appendix \ref{sec:rf_quality}. This results in identification of large number of members with high probability (here, $\sim >$ 80$\%$ members with probability $\geq$ 0.8). 
Although, the member sample retrieved so far still does not serve our purpose of investigating the sub-stellar end of the cluster, it provides a robust training sample required to augment the identification of probable cluster members down to the very low mass end.\\

% \begin{figure*}
%   \begin{subfigure*}{}
%     \begin{tabular}{|c|c|}
%      \hline
%      \multicolumn{2}{|c|}{Phase 1} \\
%      \cline{1-2}
%      Training & Feature \\
%      feature & Importance \\
%      \hline
%      pmdec & 0.25 \\ 
%      Y/r$_{2}$ & 0.21 \\
%      pmra & 0.14 \\
%      r$_{2}$ - Y & 0.13 \\
%      i$_{2}$/r$_{2}$ & 0.12 \\
%      r$_{2}$ - i$_{2}$ & 0.07 \\
%      Distance & 0.07 \\
%      \hline
%     \end{tabular}
%     %\caption{Feature $\&$ Importance}
%     %\label{tab:feat}
%   \end{subfigure*}
% %
%   \begin{subfigure*}{}
%     \begin{tabular}{|c|c|c|}
%      \hline
%      \multicolumn{2}{|c|}{Phase 2} & {Phase 3} \\
%      \cline{1-3}
%      Training & Feature & Feature \\
%      Feature & Importance & Importance \\
%      \hline
%      Y/r$_{2}$ & 0.34 & 0.36 \\
%      \hline
%      i$_{2}$/r$_{2}$ & 0.26 & 0.25 \\
%      \hline
%      r$_{2}$ - Y & 0.21 & 0.20 \\
%      \hline
%      r$_{2}$ - i$_{2}$ & 0.20 & 0.20 \\
%      \hline
%     \end{tabular}
%     \caption{{\it (Left)} Training features and their respective importance used in phase 1 of RF classifier {\it (Right)} Training features $\&$ their respective importance used in phase 2 and phase 3 of RF classifier}
%     \label{fig:feature}
%   \end{subfigure*}
  
% \end{figure*}

\begin{table}
    \centering
    \begin{tabular}{|l|l|l|l|}
     \hline
     \multirow{2}{*}{Training feature} & \multicolumn{3}{c}{Feature importance}\\
    \cline{2-4}
    & Phase 1 & Phase 2 & Phase 3\\
     \hline
     pmdec & 0.25 & - & -\\ 
     Y/r$_{2}$ & 0.21 & 0.34 & 0.36\\
     pmra & 0.14 & - & -\\
     r$_{2}$ - Y & 0.13 & 0.21 & 0.20\\
     i$_{2}$/r$_{2}$ & 0.12 & 0.26 & 0.25\\
     r$_{2}$ - i$_{2}$ & 0.07 & 0.20 & 0.20\\
     Distance & 0.07 & - & -\\
     \hline
    \end{tabular}
    \caption{Training features used in phase 1, phase 2 and phase 3 of RF classifier and their respective importance}
    \label{tab:feature}

\end{table}

We understand that the primary reason for the inability of RF classifier to identify cluster members down to the sub-stellar regime in one run is that the training sample gets biased by the colours of the brighter member stars (r$_{2}$ $\leq$ 19.5) which forms $\sim$ 50$\%$ of the training set. Due to the low signal to noise ratio, %(\citealt{2021A&A...650A..48K, 2022A&A...664A.146N, 2023ApJ...951..139D, 2023JKAS...56...97H, 2023A&A...677A..26A}), 
the scatter in optical colours increases as one approaches the fainter magnitudes. This increase in the deviation in optical colours is considered in fair estimate if we curtail the training and test set to sources with r$_{2}$ $>$ 19.5 mag for the next phase of RF classifier. The particular threshold of r$_{2}$ $>$ 19.5 mag is taken because approximately 50$\%$ of the probable members procured as of yet, are fainter than this chosen value which, provides a substantial training sample. We hence, use the training set comprising of member and non-member sample (r$_{2} >$ 19.5 mag) obtained above, and re-apply the RF classifier to the HSC sources with r$_{2} >$ 19.5 mag. With this approach, the classifier technique predicts probable members down to r$_{2} \sim$ 28 mag ($\sim$ 0.012 - 0.015 M$_{\odot}$ for an age $\sim$ 2 Myrs and A$_{V}$ $\sim$ 1 mag), which is well into the sub-stellar regime. We have provided the input features used in all the RF classifier phases along with their relative feature importance in Table \ref{tab:feature}. The quality parameters attained for each phase are also provided in Table \ref{tab:data_qual} and Figure \ref{fig: rf_qualty}. In order to be explicit, no probability threshold has been used while selecting probable members during the entire RF classifier method. Another point to note here is that the sources identified as members in all the RF classifier phases are just the probable members. %The classifier technique, predicts $\sim$ 7767 probable members down to r$_{2} \sim$ 28 mag, which is well into the sub-stellar regime.
Hence, to remove all the possible contaminants and obtain a clean and accurate list of members along with their stellar parameters like age, mass etc., we further perform SED fitting of the probable members as described in the following section.\\

%((r$_{2}$--i$_{2}$) vs r$_{2}$ ; (r$_{2}$--Y) vs r$_{2}$ and (i$_{2}$--Y) vs i$_{2}$)
%\begin{figure*}
	%\centering
	%\includegraphics[scale=0.26]{ConfusionMatrix_RF_test}
	%\includegraphics[scale=0.34]{roc1.png}
	%\includegraphics[scale=0.18]{Del_RA_spatial.png}
	%\includegraphics[scale=0.18]{Del_Dec_spatial.png}
	%\includegraphics[scale = 0.2]{cyg_centre1.eps}
	%\includegraphics[scale = 0.38]{r2-Y_vs_r2.eps}%\scriptsize
	%\includegraphics[scale = 0.4]{z-K_vs_z.eps}%\scriptsize
	%\begin{small}  
	%\scriptsize
	%\linespread{0.8}
	%\caption{ {\it left:} Confusion matrix {\it Right:} ROC curve
	 %}%{\it Top left:}
% }
%\end{small}
%\label{fig:rf}
%\end{figure*}

\subsection{VO SED Analyzer (VOSA) fitting}
\label{sec:vosa}
SED fitting is very useful to estimate the stellar parameters such as age, mass and effective temperature (T$_\mathrm{eff}$) in a star forming region (\citealt{2019ApJ...881...37P, 2020AJ....160...68W, 2021ApJ...913...45O}). %(\citealt{2007AJ....133.2072D, 2009ApJ...702..178Y, 2010A&A...518L..81Z, 2012ApJ...755...20S, 2015MNRAS.447.2307M, 2017ApJ...841..131A, 2017ApJ...840...72L, 2018A&A...619A..52B, 2019MNRAS.489.2615M, 2019ApJ...881...37P, 2020AJ....160...68W, 2021ApJ...909..121M, 2021ApJ...913...45O}). 
The determination of these parameters helps to understand the evolution of the stellar cluster with respect to age, mass and environmental factors. %, thus implying its habitability for planet formation (\citealt{2014MNRAS.438.2413V, 2015ApJ...800..115T, 2018AJ....155...22S, 2019ApJ...886..131G, 2021A&A...646A..46G, 2022MNRAS.510.3449B, 2023AJ....165..155T, 2023A&A...671A.140G}). 
In order to obtain a precise estimation of these stellar parameters, a broad wavelength range for the spectral fitting is favourable. Hence, we use UKIDSS DR11/ 2MASS NIR data along with the deep optical data for this purpose. Among the RF classifier identified probable members, we consider only those 1156 HSC sources, which have counterparts in the deep NIR catalog.\\
We use the VOSA\footnote{http://svo2.cab.inta-csic.es/theory/vosa/} tool to build the SEDs using optical to NIR wavelengths (\citealt{2008A&A...492..277B, 2021MNRAS.504.2557D, 2021A&A...650A.182G, 2022A&A...660A.131M}). However as per the requirements of VOSA, we consider only those sources for SED fitting which have photometry in atleast five filters. In addition to the provided photometry, VOSA also uses Pan-STARRS photometry from the online catalogues whenever present using VO services. We also input a distance (900 $\pm$ 10 parsecs) as well as the extinction range (A$_{V}$ $\sim$ 1 - 1.5 mag) estimates in accordance with the past studies (see references mentioned in Section \ref{sec: IC 1396}), to deredden the SEDs. Subsequently, theoretical models (BT-Settl, \citet{2012RSPTA.370.2765A}) for solar metallicity are used to obtain the best fitting model using chi-square minimization. This gives us the physical properties like T$_\mathrm{eff}$ and luminosity of the  stars. The age and mass is derived by fitting the sources in the Hertzsprung-Russell (HR) diagram with BHAC15 isochrones (\citealt{2015A&A...577A..42B}) for a mass range 0.01 - 1.4 M$_{\odot}$ and age ranging between 0.5 - 10 Myrs. Those sources which do not fulfill these constraints are rejected as contaminants. %The fitting process uses input parameters like T$_\mathrm{eff}$, metallicity and surface gravity (logg) for the model.
%The default values of input parameters, i.e T$_\mathrm{eff}$ $\sim$ 1200 - 7000 K, metallicity = 0 and logg $\sim$ 2.5 - 5.5 are used during the SED fitting process. After fitting SED with the models we obtain the T$_\mathrm{eff}$ and luminosity of the individual sources. This when fitted with BHAC15 isochrones (\citealt{2015A&A...577A..42B}) in the Hertzsprung-Russell (HR) diagram, finally provides information about age and mass of the sources. The BHAC15 isochrones are available for a mass range of 0.01 - 1.4 M$_{\odot}$ and age ranging between 0.5 - 10$^{4}$ Myrs. The mass and age information may be missing for field stars or variable stars which do not lie within these constraints. Hence, such non-members lack the fitting by models shifted for the distance (900 parsecs) and Av (1 mag) of this cluster, appropriately. We find that the respective mass and age information is absent for $\sim$ 40$\%$ of the total 2078 sources which are hence, rejected here. We only select sources with available age and mass as members, along with the prerequisite that evaluated age be $\leq$ 10 Myrs. 
At this stage, we also remove those sources with Gaia DR3 counterparts which do not fulfil the constraints mentioned in Section \ref{sec:gmm} and Section \ref{sec:gaia data} (for example, 500 parsecs $\leq$ distance $\leq$ 1500 parsecs, ruwe $<$ 1.4, --20 $\leq$ $\mu_{\alpha}$cos$\delta$ $\leq$ 20 mas/yr , --20 $\leq$ $\mu_{\delta}$ $\leq$ 20 mas/yr). This finally gives us 393 sources as confirmed member sources within the target region of study using machine-learning methods.\\

We summarise the machine-learning based approach used to identify the cluster members within the central 22$^{\prime}$ radius region here. We first obtain a Gaia-based member sample with probability $\geq$ 0.8 using GMM. This sample is then used as the training set in RF classifier to identify cluster members among the sources with both HSC and Gaia photometry. The members identified at this step are further used as training sample for the next phases of RF classifier to segregate members among the HSC sources without Gaia counterparts. The members obtained at each step are compiled together and cross-matched with the deep NIR catalog. SED fitting of these probable members with both optical and NIR photometry is performed using VOSA. We remove sources with Gaia counterparts which do not comply with membership constraints. We finally select 393 sources with age $<$ 10 Myrs, as cluster members. The identified members however suffer from the restricted sensitivity of UKIDSS (complete down to $\sim$ 0.09 M$_{\odot}$) due to the pre-requisite of having NIR counterparts which is essential for SED fitting. Hence, we adopt an additional approach to identify fainter cluster members by defining a pre-main sequence locus as explained below in Section \ref{sec:pms locus}.\\  % deep optical data (that includes, HSC and Pan-STARRS photometry). Although these sources have mass ranging down to 0.02 M$_{\odot}$, the completeness is poor towards the fainter end. Since UKIDSS data is 90$\%$ complete down to J $\sim$ 16 mag which corresponds to $\sim$ 0.09 M$_{\odot}$ (r$_{2}$ $\sim$ 22 mag) for the cluster age $\sim$ 2 Myrs. In the above method based on SED, many probable candidates would have been missed due to UKIDSS sensitvity. Thus, it is crucial to extend the member identification to deeper sensitivity independent of NIR counterparts.

\subsection{Additional members from pre-main sequence locus}
\label{sec:pms locus}
\begin{figure*}
	%\centering
	%\includegraphics[width=8.5cm, height=6cm]{i2Y_cmd_Yso.png}
	\includegraphics[scale=0.27]{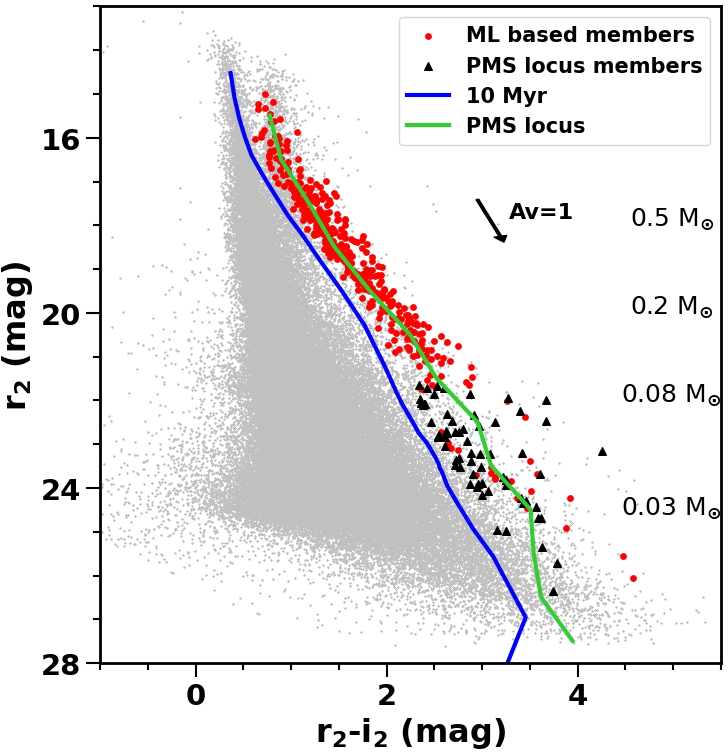}
	\includegraphics[scale=0.27]{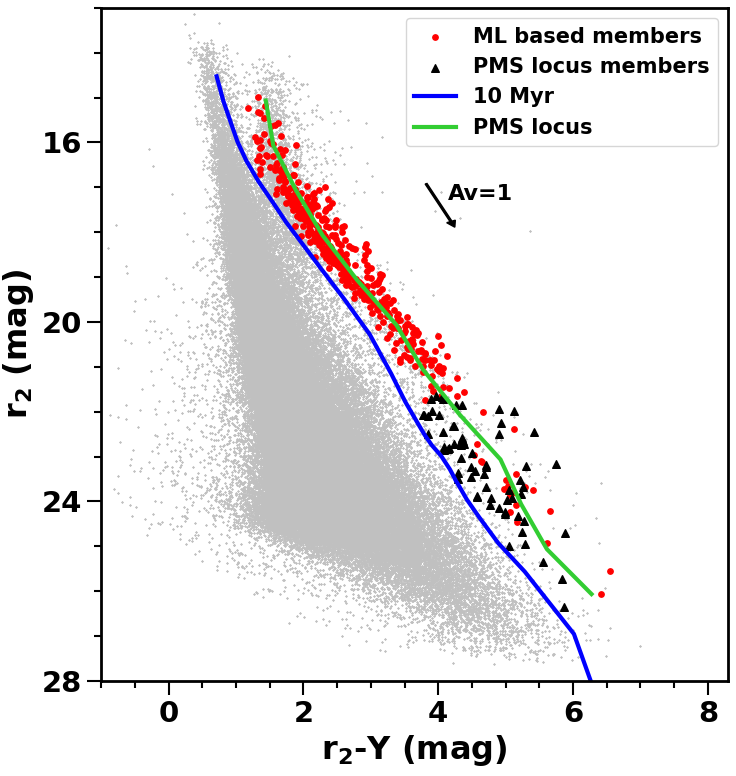}
	%\includegraphics[scale = 0.38]{r2-Y_vs_r2.eps}%\scriptsize
	%\includegraphics[scale = 0.4]{z-K_vs_z.eps}%\scriptsize
	%\begin{small}  
	%\scriptsize
	%\linespread{0.8}
	\caption{({\it Left}) r$_{2}$ - i$_{2}$ vs r$_{2}$ colour-magnitude ({\it Right}) r$_{2}$ - Y vs r$_{2}$ colour-magnitude diagram to obtain members using pre-main sequence locus (green curve). Blue curve shows the 10 Myrs BHAC15 isochrone. The newly added members are presented as black triangles.
	 } 

%{\it Top left:}
% }
%\end{small}
\label{fig: PMS_cmd}
\end{figure*}

It is crucial to extend the member identification to deeper sensitivity independent of NIR counterparts, to perform a comprehensive study of the low-mass population in IC 1396. In order to detect lower mass cluster members ($<$ 0.09 M$_{\odot}$), we define the respective empirical pre-main sequence (PMS) locus for all optical colour-magnitude combinations ((r$_{2}$ - i$_{2}$) vs r$_{2}$, (r$_{2}$ - Y) vs r$_{2}$ and (i$_{2}$ - Y) vs i$_{2}$) by dividing the magnitude range into 1 mag bins. For each bin, we take mean of the magnitude and median of the colour of sources inside the bin (\citealt{2021MNRAS.504.2557D, 2021MNRAS.508.3388G}). We then select only those sources which are fainter than r$_{2}$ $=$ 22 mag and within 1 $\sigma$ (where $\sigma$ is the standard deviation) limits of the empirical locus defined for the respective colour-magnitude combination. Also, the sources are required to be common in all the colour-magnitude combinations, which adds to the validation of these sources as cluster members. The particular threshold of r$_{2}$ $>$ 22 mag is chosen to select the members here as this corresponds to a stellar mass less than $\sim$ 0.09 M$_{\odot}$ (for cluster age and distance). This corresponds to the data completeness of UKIDSS data which restricts the sensitivity of ML-based members obtained in Section \ref{sec:vosa}. We obtain 65 members with this approach, thus increasing the total number of cluster members to 458 and detection limit down to 0.03 M$_{\odot}$.

Figure \ref{fig: PMS_cmd} shows r$_{2}$ - i$_{2}$ vs r$_{2}$ and r$_{2}$ - Y vs r$_{2}$ colour-magnitude diagrams overplotted with ML-based, pre-main sequence locus based cluster members, the defined pre-main sequence locus and 10 Myrs BHAC15 isochrone (\citealt{2015A&A...577A..42B}). The BHAC15 isochrone is corrected for an extinction A$_V$ $\sim$ 1 mag and distance $\sim$ 900 parsecs. We use the \citet{2019ApJ...877..116W} empirical laws to correct for the extinction. Since BHAC15 isochrones are not available for Subaru HSC photometric system, we first convert the available Pan-STARRS 10 Myrs isochrone to HSC system using the transformation equations given in \citet{2021MNRAS.508.3388G}. Subsequently, we correct the 10 Myr isochrone for the distance and reddening of the cluster. In order to determine the mass, these 65 members are first de-reddened for the uniform cluster reddening and age. The mass for each source is then determined using the mass - magnitude relation obtained using BHAC15 models. This member sample is useful to identify brown dwarfs and obtain the mass distribution in the region. We would like to mention here that although, reddening across the central 22$^{\prime}$ radius region of IC 1396 is uniform with A$_{V}$ $\sim$ 1 - 1.5 mag, we check if the variation in reddening values may change the output member sample significantly. Hence, we verify the error introduced by varying the A$_{V}$ upto 1.5 mag and find that a minimal error of $\sim$ 2$\%$ is introduced in the statistics of cluster members. Since, the output is approximately uniform, we continue using A$_{V}$ $=$ 1 for our analysis.\\ %We caution the readers that although we have been conservative in our approach to select cluster members with this method, minor contamination in the sample is probable. These contaminants however can not be removed as it requires spectroscopy of the individual stars which is out of the scope of this study.\\

In order to verify the validity of member sample obtained by this approach, we perform field subtraction of the central 10$^{\prime}$ radius region of IC 1396. For this purpose, we choose a control field of the same area towards the periphery of the region (HSC data for this peripheral region is available by the wide field observations of IC 1396 as mentioned in Section \ref{sec:data}). The 10$^{\prime}$ radius region is used due to the lack of a larger control field. The statistical field subtraction is then performed, the details of which are provided in Appendix \ref{sec:fs}. The same approach as described above in this section is then used to select members among these statistically field decontaminated sources. We find that the statistics of members and corresponding mass distribution obtained in Section \ref{sec: imf} remain approximately same whether or not prior field subtraction is performed. This uniformity hence, validates the approach adopted by us above. We therefore, continue with the members obtained above without using field decontamination for further analysis in the paper (detailed supporting reasons are explained in Appendix \ref{sec:fs}).\\

\subsection{Comparison with previous studies}
\label{sec: literature}
The identified cluster members are high-quality sources with photometric error $\leq$ 0.1 in optical filters. The proper motions and parallax range of the obtained sources with Gaia counterparts ($\sim$ 85$\%$) are in excellent accordance with that for literature based sources (\citealt{2002AJ....124.1585C, 2006ApJ...638..897S, 2007ApJ...654..316G, 2012MNRAS.426.2917G, 2013A&A...559A...3S, 2018A&A...618A..93C, 2023A&A...669A..22P, 2023ApJ...948....7D}), as shown in Figure \ref{fig: gaia_qualty} {\it Top} and {\it Bottom}. The central 22$^\prime$ radius area of IC 1396 comprises of 857 members ascertained by previous studies such as those mentioned above. An additional $\sim$ 330 members have been identified by the recent Gaia based studies like \citet{2018A&A...618A..93C, 2022arXiv221011930P} and \citet{2023ApJ...948....7D} within the area of study. Out of the total 458 member sources confirmed in our study, we retrieve 328 members with counterparts in the literature studies (including both previous studies and recent Gaia-based studies mentioned above). The remaining 130 members without counterparts in literature are low mass stars with r$_{2}$ $\geq$ 18 mag, that is, mass $<$ 0.5 M$_{\odot}$. This low and sub-stellar mass range is still unexplored for IC 1396 which explains the absence of literature counterparts of these faint sources. %Approximate $\sim$ 60$\%$ -- 65$\%$ of the members procured by recent studies such as \citet{2018A&A...618A..93C, 2022arXiv221011930P, 2023ApJ...948....7D} have counterparts with our member list. The remaining 35$\%$ - 40$\%$ sources which do not find counterparts in our study, are located either towards the brighter end (r$_{2}$ $<$ 18 mag) or scattered to the left of 10 Myrs isochrone and hence, rejected in our analysis due to the 10 Myr condition imposed in our selection. %, but the low fraction of counterparts with other literature studies like \citet{2004ApJS..154..385R, 2006ApJ...638..897S, 2007ApJ...654..316G, 2009ApJ...702.1507M, 2013A&A...559A...3S, 2021AJ....162..279S}, is primarily due to different fields centered towards the periphery or out of our area of study.  

\begin{figure}
	%\centering
	%\includegraphics[scale=0.5][trim={0 0 2 2},clip]{HRD_hess.png} \adj ,trim={0 0 {.12\width} {0.1\height} 
    \includegraphics[scale=0.27]{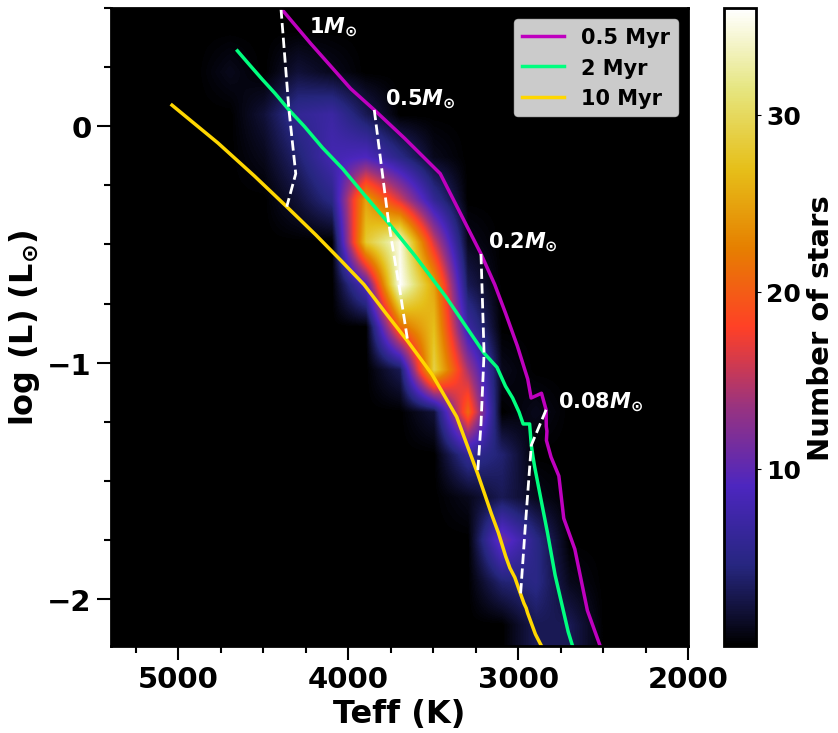}
	%\includegraphics[width=8.6cm, height=6cm]{Del_Dec_spatial.png}
	%\includegraphics[scale=0.18]{Del_RA_spatial.png}
	%\includegraphics[scale=0.18]{Del_Dec_spatial.png}
	%\includegraphics[scale = 0.2]{cyg_centre1.eps}
	%\includegraphics[scale = 0.38]{r2-Y_vs_r2.eps}%\scriptsize
	%\includegraphics[scale = 0.4]{z-K_vs_z.eps}%\scriptsize
	%\begin{small}  
	%\scriptsize
	%\linespread{0.8}
	\caption{ Hess plot of HR diagram of the members obtained from VOSA. BHAC15 isochrones of 0.5, 2 and 10 Myr are over-plotted along with mass evolutionary tracks as white dashed curves.
	 }%
% }
%\end{small}
\label{fig:hrd}
\end{figure}

Our study adds 130 new members to the existing cluster member catalog with detection limit down to $\sim$ 26.5 mag in r$_{2}$-band, which corresponds to a $\sim$ 0.025 M$_\odot$ source for the distance and age of the cluster. This significantly enhances the sensitivity down to where IC 1396 can be studied in future. The Hess plot of the HR diagram of members identified in the present study is presented in Figure \ref{fig:hrd}. We use this member list for further analysis in this paper.

%\textbf{Our study adds $\sim$ 121 new members to the existing cluster member catalog, out of which $\sim$ 90$\%$ of the member stars have r$_{2} \geq$ 19 mag}. This corresponds to a stellar mass $\leq$ 0.3 M$_{\odot}$ which, significantly enhances the sensitivity down to where IC 1396 can be studied in future. % Although the identified members are limited by the UKIDSS sensitivity, 
%With the identified cluster members, we are able to investigate the faint sub-stellar regime down to $\sim$ 26.5 mag in r$_{2}$-band, which corresponds to a $\sim$ 0.025 M$_\odot$ source for the distance and age of the cluster. The Hess plot of the HR diagram of members identified in the present study is presented in Figure \ref{fig:hrd}. We use this member list for further analysis in this paper.\\ %Approximately, 196 of the 214 sources (limiting magnitude, r$_{2}$ $\sim$ 26 mag) (complete down to 24.5 mag, corresponding to $\sim$ 0.03 M$_\odot$)

\section{Results and Analysis}
\label{sec:analysis}

\subsection{Sub-stellar members and their characteristics}
\label{sec:sub-stellar}
%\begin{figure}
	%\centering
	%\includegraphics[scale=0.24]{age_hist.png}
	%\caption{ Age histogram for all the identified members. The median age of the cluster members is found to be log (Age) $\sim$ 6.3 $\pm$ 0.6, which in linear scale implies 2 $\pm$ 1.5 Myrs. 
%	 }
	 
%{\it Top left:}
% }
%\end{small}
%\label{fig:age_whole}
%\end{figure}
We use the obtained members to perform statistical analysis of the region. We determine the sub-stellar population and consequently determine the mass distribution within the studied area. We use the obtained stellar masses of individual member sources in Section \ref{sec:vosa} and Section \ref{sec:pms locus} to determine if they can be classified as a sub-stellar object. We identify 80 sources with mass $\le$ 0.1 M$_\odot$ as sub-stellar objects. Out of these, 62 sources have mass $\leq$ 0.08 M$_\odot$, which we classify as brown dwarfs. %The brown dwarfs  in the studied 22$^\prime$ region have T$_\mathrm{eff}$ $\leq$ 2900 K which confirms the nature of these cool objects (\citealt{2010ApJS..186...63R, 2015ApJ...810..158F}).
We also derive the age of the member sample using isochrone fitting method. In this method, for each star a weighted average of the age of two nearest isochrones is assigned as the age of the star. We converge this distribution within 2$\sigma$ limits from the mean age after 3 iterations (\citealt{2021MNRAS.508.3388G}). We find the median age of the 2$\sigma$ converged sample to be log (Age) $\sim$ 6.3 $^{+0.2}_{-0.5}$, which complies with the age of the IC 1396 cluster as estimated in literature (\citealt{2005AJ....130..188S, 2012MNRAS.426.2917G, 2012AJ....143...61N, 2023ApJ...948....7D}).\\

\subsection{Colour-magnitude and Colour-colour diagrams}
\label{sec:cmd}

\begin{figure*}
	%\centering
	%\includegraphics[width=8.5cm, height=6cm]{r2z_cmd_whole.png}
%	\includegraphics[width=8.5cm, height=6cm]{r2z_cmd_whole.png}
	\includegraphics[scale=0.212]{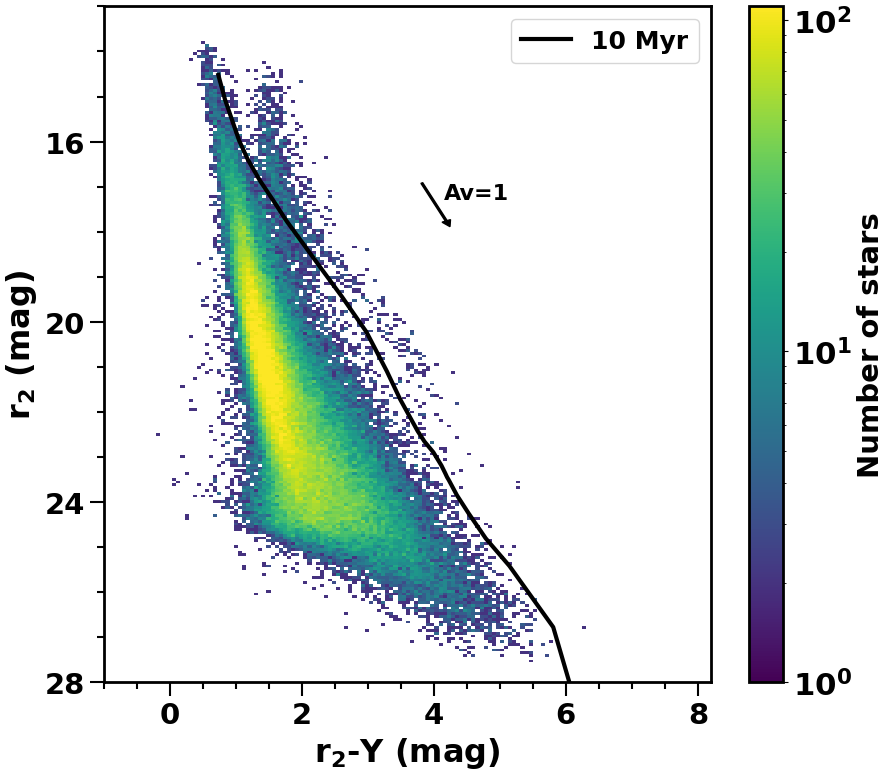}
	\includegraphics[scale=0.2138]{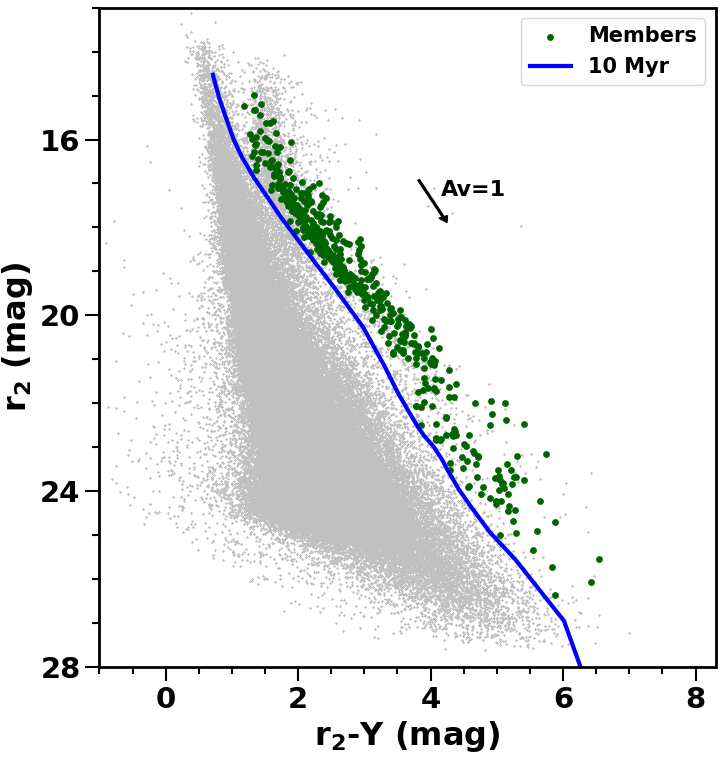}
	\includegraphics[scale = 0.2143]{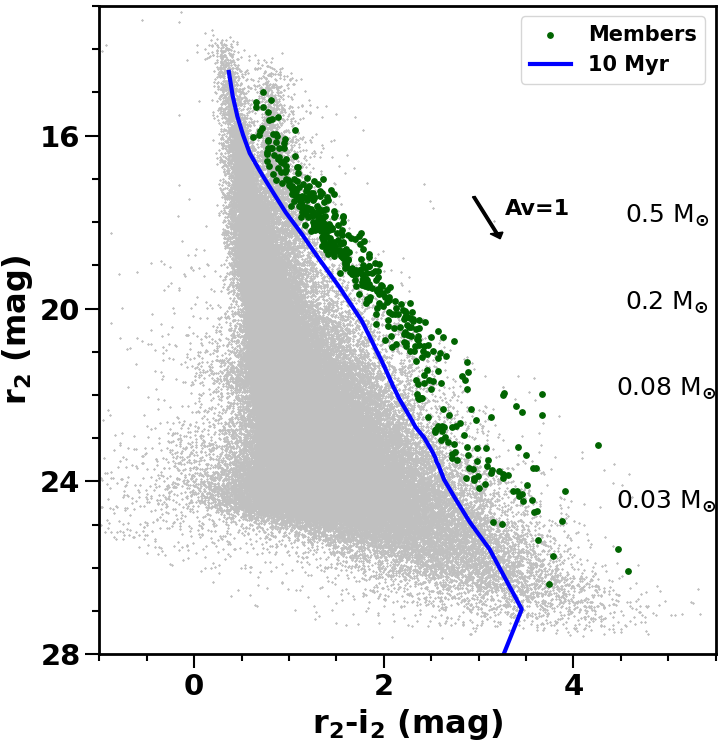}%\scriptsize
	%\includegraphics[scale = 0.4]{z-K_vs_z.eps}%\scriptsize
	%\begin{small}  
	%\scriptsize
	%\linespread{0.8}
	%\caption{ colour-Magnitude plots
	 %} 
	 
%{\it Top left:}
% }
%\end{small}
%\label{fig:cmd}
\end{figure*}

\begin{figure*}
	%\centering
	%\includegraphics[width=8.5cm, height=6cm]{r2z_cmd_whole.png}
%	\includegraphics[width=8.5cm, height=6cm]{r2z_cmd_whole.png}
	\includegraphics[scale=0.215]{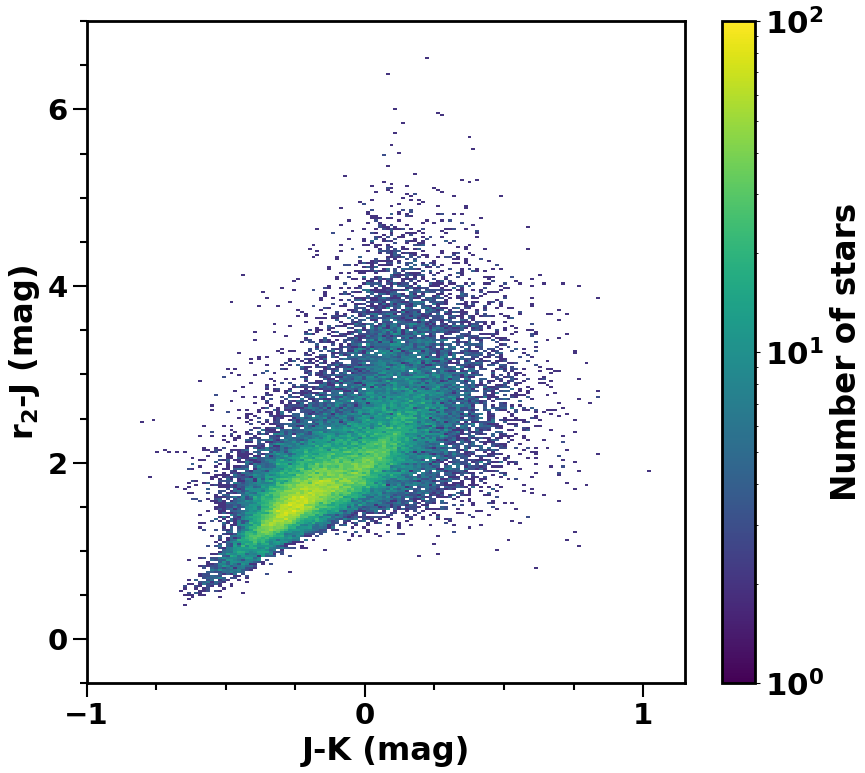}
	\includegraphics[scale=0.225]{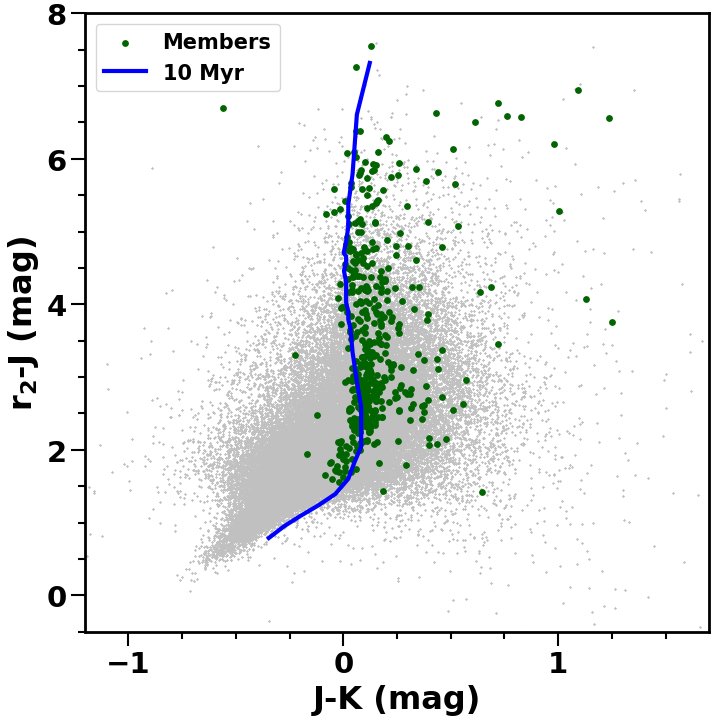}
	%\includegraphics[scale = 0.4]{z-K_vs_z.eps}%\scriptsize
	%\begin{small}  
	%\scriptsize
	%\linespread{0.8}
	\caption{ ({\it Top Left}) Hess plot of (r$_{2}$--Y) vs r$_{2}$ CMD ({\it Top Center}) (r$_{2}$--Y) vs r$_{2}$ CMD scatter plot ({\it Top Right}) (r$_{2}$--i$_{2}$) vs r$_{2}$ CMD scatter plot ({\it Bottom Left}) Hess plot of (r$_{2}$--J) vs J--K colour-colour plots ({\it Bottom Right}) (r$_{2}$--J) vs J--K scatter colour-colour plots overplotted with 10 Myrs BHAC15 isochrone(\citealt{2015A&A...577A..42B}) and identified members (dark green dots).
	 } 
	 
%{\it Top left:}
% }
%\end{small}
\label{fig:cmd_ccd}
\end{figure*}

We present here the various colour-magnitude and two-colour diagrams for the 22$^\prime$ radius area of study in IC 1396. Both the Hess plots of (r$_{2}$--Y) vs r$_{2}$ CMD and (r$_{2}$--J) vs (J--K) two colour plot (Figure \ref{fig:cmd_ccd} {\it (Top Left)} and {\it (Bottom Left)}) exhibit a distinct pre-main sequence branch. The corresponding scatter plots (Figure \ref{fig:cmd_ccd} ({\it Top Center}, {\it Top Right} and {\it Bottom Right})) are overplotted with the identified members and 10 Myr BHAC15 isochrone corrected for the cluster extinction A$_V$ $\sim$ 1 mag and distance $\sim$ 900 parsecs. %We use the \citet{2019ApJ...877..116W} empirical laws to correct for the extinction. Since BHAC15 isochrones are not available for Subaru HSC photometric system, we first convert the available Pan-STARRS 10 Myrs isochrones to HSC system using the transformation equations given in \citet{2021MNRAS.508.3388G}. Subsequently, we correct the 10 Myr isochrone for distance and reddening using above mentioned distance and extinction values.
In these plots, we observe that the identified stellar and sub-stellar members in the region pre-dominantly occupy the distinct pre-main sequence branch, which is a typical feature of young star forming regions (\citealt{2017ApJ...836...98J, 2019A&A...623A.112D, 2021MNRAS.504.2557D, 2021MNRAS.508.3388G, 2021AJ....161..257K}).\\

\subsection{Mass distribution}
\label{sec: imf}
\begin{figure}
	%\centering
	%\includegraphics[width=8.5cm, height=6cm]{i2Y_cmd_Yso.png}
	\includegraphics[scale=0.23]{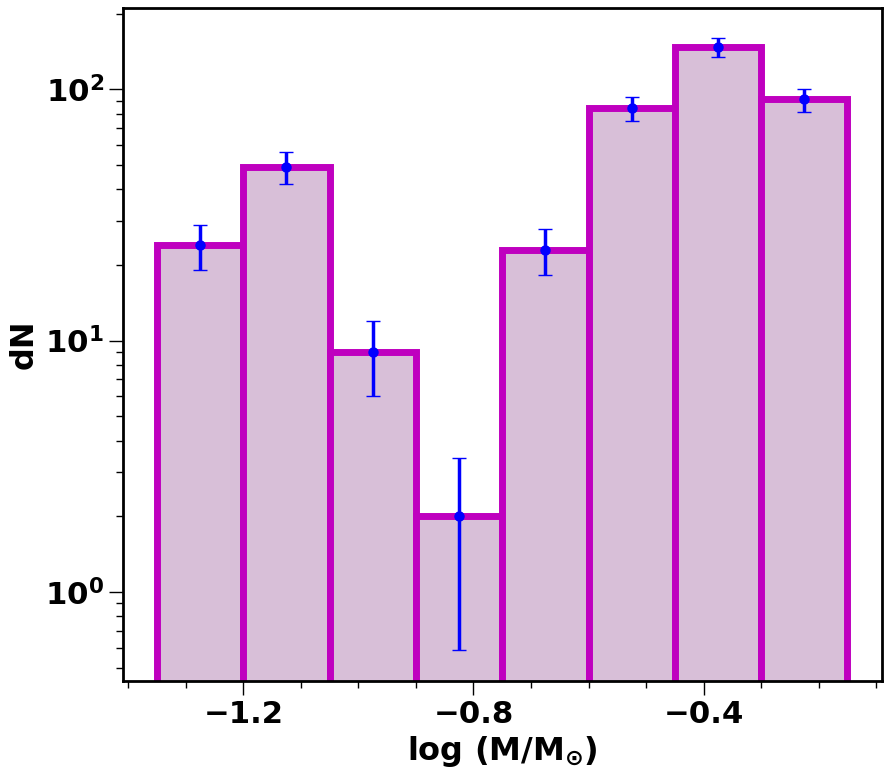}
	%\includegraphics[width=6.2cm, height=9cm]{ageBin_hist.png}
	%\includegraphics[scale = 0.38]{r2-Y_vs_r2.eps}%\scriptsize
	%\includegraphics[scale = 0.4]{z-K_vs_z.eps}%\scriptsize
	%\begin{small}  
	%\scriptsize
	%\linespread{0.8}
	\caption{ Mass distribution plot for the identified cluster members. The primary peak is $\sim$ 0.4 M$_{\odot}$. A secondary peak is observed at $\sim$ 0.06 M$_{\odot}$ and a dip is observed at $\sim$ 0.1 - 0.15 M$_{\odot}$. The error-bars represent the Poisson error in the count of sources in each mass bin.
	 } 

%{\it Top left:}
% }
%\end{small}
\label{fig:imf}
\end{figure}

The distribution of stellar mass in a cluster is crucial to understand as well as constrain the star formation process (\citealt{2012ApJ...753..156K, 2014prpl.conf...53O, 2021ApJS..253....7K, 2023arXiv231203639K}). The obtained member population provides a fair sample of low-mass sources to investigate the nature of mass distribution towards the sub-stellar end. It is also helpful to check the role played by the feedback regulated cluster environment of IC 1396 on the mass apportionment in the region. Due to data completeness issues, we consider the members within the mass range 0.03 -- 1 M$_\odot$ for our analysis. % since  the r$_{2}$ band photometry is complete down to $\sim$ 0.03 M$_{\odot}$ as mentioned above in Section \ref{sec:data}.
%The IMF is described by several formalisms including the power-law form (\citealt{1955ApJ...121..161S}), segmented power-law form (\citealt{2001MNRAS.322..231K}) and the logarithmic form (\citealt{2003PASP..115..763C}). We find that the single power-law form fits best to the obtained mass distribution.\\
%\begin{equation}
%    \begin{split}
%    \phi(log m) = \frac{dN}{d  log m} & \propto m^{-\Gamma}\\
%    %\xi(logm) & \propto e^{{-}\frac{(logm-logm_{c})^2}{2\sigma^2}}\\    
%    \end{split}
%\end{equation}
%where, N is the number of stars per logarithmic mass bin and $\Gamma$ is the slope of this distribution.\\ 
The distribution of stellar mass among the cluster members is estimated by dividing the entire logarithmic mass range into bins of size, log M $=$ 0.15 and the corresponding number of sources in each bin is calculated. The corresponding histogram is shown in Figure \ref{fig:imf}. The Poisson error in the count of sources is represented by the error-bars in each mass bin of the histogram. We observe a bimodal nature of mass distribution with a secondary peak at $\sim$ 0.06 M$_\odot$ (log (M/M$_\odot$) $\sim$ -1.2) as the distribution advances in sub-stellar regime. We also observe a sharp dip in the distribution at $\sim$ 0.10 - 0.15 M$_\odot$. The bimodal nature of the mass distribution with a secondary peak at $\sim$ 0.06 M$_{\odot}$ and the pronounced dip at the stellar-sub-stellar boundary agrees well with that observed for ONC by \citet{2016MNRAS.461.1734D}. The authors observed a similar bimodal nature of IMF in the feedback-driven ONC region. This suggests a similar brown dwarf formation scenario in both IC 1396 and ONC with similar feedback affected cluster environments. A similar secondary peak has been observed in NGC 2024 by \citet{2006ApJ...646.1215L} and Trapezium cluster by \citet{2004ApJ...610.1045S} at $\sim$ 0.03 - 0.05 M$_{\odot}$. However, uncertainties in the derived age of sources affects the mass-luminosity relation, which may influence shape of mass distribution.  %This implies a deficit of brown dwarfs compared to the stellar population in the region.  The rapid decline of the mass distribution towards brown dwarf regime is in accordance with the downturn in the sub-stellar mass function theorized by \citet{2001MNRAS.322..231K, 2005ASSL..327...41C}. We can easily estimate the linear slope as ${\alpha}$ = 0.6, using (\citealt{2010ARA&A..48..339B}):
%The single power-law slope $\alpha$ has been found to range between 0.4 -- 1 for several of the star forming regions like $\rho$ Oph, $\epsilon$ Cha, NGC 2264, NGC 1333, NGC 2244 (\citealt{2012ApJ...744..134M, 2012ApJ...744....6S, 2013A&A...549A.123A, 2021MNRAS.504.2557D, 2021A&A...650A..48K, 2021MNRAS.507.4074P, 2023ApJ...951..139D, 2023A&A...677A..26A}). The obtained IMF slope is hence, compatible with that obtained for other star forming regions within the error range. It is observed that the IMF for a cluster is independent of the cluster environment. For example, $\alpha \sim$ 0.3 and 0.7 for high feedback impacted regions like Westerlund 1 and RCW 38 respectively (\citealt{2017A&A...602A..22A, 2019ApJ...881...79M}). Similarly for low feedback affected regions like Lupus 3, Cha I $\alpha \sim$ 0.7 - 0.8 (\citealt{Mu_i__2015}) which, is well within the standard range for power-law slopes.  
%However, we caution our readers that since the completeness of our members is limited by UKIDSS sensitivity, this is the lower limit of the number of sub-stellar members found in the region. We also observe that a single power law fit extends continually down to the sub-stellar realm ($\leq$ 0.1 M$_\odot$) which suggests a common mechanism of formation of both sub-stellar and stellar sources. We discuss and interpret these results further in Section \ref{sec:discuss}. 
The current study pioneers to explore mass distribution down to the so far unexplored sub-stellar realm in the central 22$^\prime$ radius region of IC 1396. %A previous study of Tr 37 by \citet{2013AN....334..673E} uses XMM X-ray data to obtain the IMF fitted with an uncommon power law index $\sim$ 1.9 $\pm$ 0.44 for a mass range of 0.1 -- 0.5 M$_\odot$. However, the authors suggest a more deeper study of the region for any conclusive result.\\
%(Kirkpatrick et al. 2012, Muzic et al. 2017, Huston $\&$ Luhman 2021)

\subsection{Star-to-Brown Dwarf Ratio}
\label{sec:sbd ratio}

\begin{figure}
	%\centering
	%\includegraphics[width=8.5cm, height=6cm]{i2Y_cmd_Yso.png}
	\includegraphics[scale=0.2]{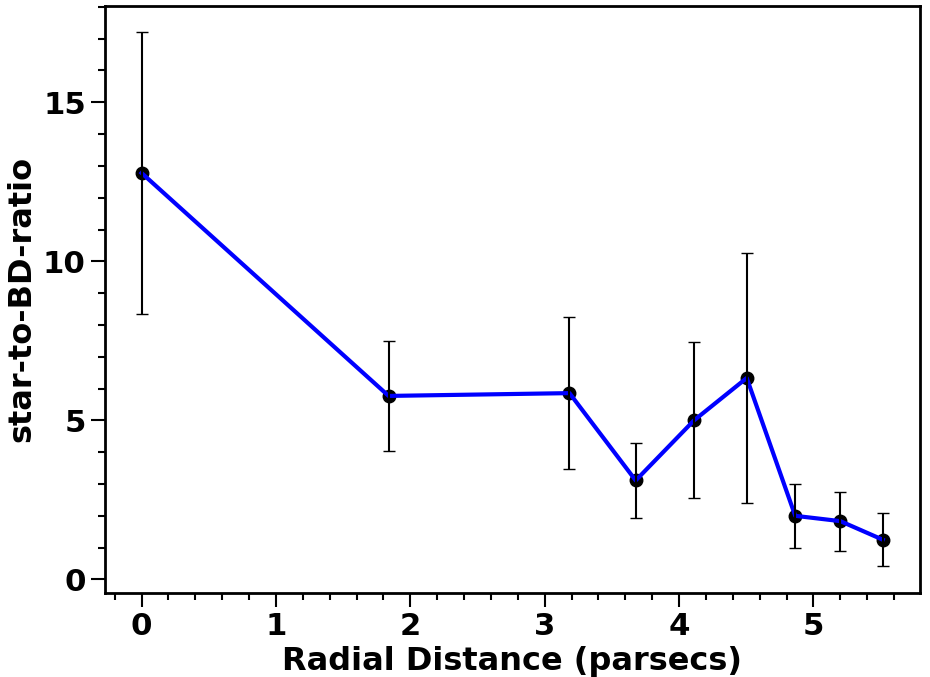}
	%\includegraphics[width=6.2cm, height=9cm]{ageBin_hist.png}
	%\includegraphics[scale = 0.38]{r2-Y_vs_r2.eps}%\scriptsize
	%\includegraphics[scale = 0.4]{z-K_vs_z.eps}%\scriptsize
	%\begin{small}  
	%\scriptsize
	%\linespread{0.8}
	\caption{ Variation of the star-to-brown dwarf ratio with respect to distance from the central OB-stars (HD 206267), across the 22$^{\prime}$ radius area of study in IC 1396. Each point refers to the star-to-brown dwarf ratio in individual annular bin and the corresponding Poisson error represented as error bar.
	 } 

%{\it Top left:}
% }
%\end{small}
\label{fig:radial_sbd}
\end{figure}

Another useful quantity which gives a fair estimate of the abundance of brown dwarfs in a region is the star-to-brown dwarf ratio R. A high R value indicates lower brown dwarf fraction in a stellar cluster. The ratio thus implies the brown dwarf formation efficiency in a cluster. An objective comparison of the ratio across varying star forming environments is important to understand if the brown dwarf formation is impacted by the external factors (\citealt{Mu_i__2015, 2017MNRAS.471.3699M, 2019ApJ...881...79M, 2021AJ....161..138H, 2023ApJ...951..139D, 2023A&A...677A..26A}). We consider brown dwarfs only within 0.03 - 0.08 M$_{\odot}$ mass range for the R value calculation, since the completeness of the data is down to 0.03 M$_{\odot}$. %This difference in the count of brown dwarfs when extrapolated is due to data incompleteness as cautioned in Section \ref{sec: imf}.
With the count of stars ranging within 0.08 - 1 M$_{\odot}$ estimated as 391, the star-to-brown dwarf ratio in IC 1396 turns out to be $\sim$ 6.0 $\pm$ 0.8, i.e $\sim$ 6 stars for each brown dwarf in the region. The error in R is calculated as the Poisson error for the ratio. We emphasize that due to the completeness issues mentioned above, the obtained R value is evaluated for the stellar mass range 0.03 - 1 M$_{\odot}$. The upper limit of stellar mass to determine R is taken as 1 M$_{\odot}$ so as to make a fair comparison between IC 1396 and other star forming regions. We observe that although towards the higher end, the obtained ratio R is compatible with other young star forming regions within the statistical uncertainties for the mass range 0.03 - 1 M$_{\odot}$. For example, the ratio R ranges within 2 - 6 in literature for several star forming regions with diverse cluster environments like $\rho$ Oph, IC348, Lupus3, NGC 2264, NGC 2244 (\citealt{2012ApJ...744..134M, 2013ApJ...775..138S, Mu_i__2015, 2023ApJ...951..139D, 2023A&A...677A..26A, 2023arXiv231203639K}) as given in Table \ref{tab:sbd_den_uv}. This implies a similar brown dwarf formation process across various star forming regions in the Galaxy.

In addition, we check the variation of ratio R with increasing radial distance from the central O-star HD 206267, as shown in Figure \ref{fig:radial_sbd}. This is obtained by dividing the 22$^{\prime}$ radius region into annular bins of equal area, here taken to be 50$\pi ^{\prime}$$^{2}$. The star-to-brown dwarf ratio is then calculated for each annular bin based on the number of stars and brown dwarfs present in it. The error bars at each point represent the Poisson error obtained for the ratio R in each bin. The limited census of brown dwarfs account for the high error observed in the ratio R in the inner annular bins towards the center of the cluster. The ratio R is observed to decrease as one moves radially outwards away from the center. This implies that more brown dwarfs are located towards the outskirts of the studied area of interest compared to its central part where the ionizing system HD 206267 is located. The obtained R ($\sim$ 6) averaged over our area of study is clearly biased by the very low brown dwarf fraction (and hence, considerably higher R values) towards the inner radii ($<$ 3 parsecs, $\sim$ 12$^{\prime}$) of IC 1396. This is likely contributed by the O-stars at the center of the cluster. As evident from Figure \ref{fig:radial_sbd}, the ratio R falls to lower values at distances $>$ 3 parsecs from the center. The presence of O-stars thus, seems to inhibit the formation of brown dwarfs towards the center of IC 1396. The brown dwarf fraction increases as the incident FUV flux from the central massive stars decreases across the region. It is hence evident that the feedback driven cluster environment of IC 1396 affects the spatial distribution of brown dwarfs within the region.\\ %in the region may be supported by the premature ejection of low-mass cores. Such self-gravitating cores are cut-off from their material reservoir and expelled towards the outer radii thus, resulting into a high fraction of brown dwarfs towards the outskirts of the studied area. However, this requires a comprehensive investigation of the kinematics of low-mass and sub-stellar sources for a larger area of IC 1396, which is out of the scope of this study.}\\

\section{Discussion}
\label{sec:discuss}

The cluster environment in which stars are born plays a crucial role in constraining the star formation and regulating the timescale for related processes like brown dwarf formation, disk evolution and planet formation. The quantitative effect of environmental factors on these processes is an interesting problem yet to be fully understood. Although many studies (\citealt{2010ARA&A..48..339B, 2014prpl.conf...53O, 2021MNRAS.504.2557D}) suggest a uniformity in IMF behaviour in the high-mass end and hence, a similar star formation process across various star forming regions, but a non-uniform IMF has been claimed in extreme environments like the Galactic centre and least luminous Milky Way satellites (\citealt{2013JGRA..118.3113L, 2018ApJ...855...20G}). Similarly, various theoretical studies also suggest an enhancement in the census of sub-stellar objects in the presence of high gas and stellar density or in the vicinity of massive stars (\citealt{2004A&A...427..299W, 2012MNRAS.419.3115B, 2013MmSAI..84..866V, 2014ASSP...36...17S}). However, more observational evidences in different star forming environments are still required to support results from theoretical simulations. In this section we interpret our results and compare them with other star forming regions as well.\\

We observe from our study that the mass distribution has a secondary peak in the brown dwarf regime (log (M/M$_{\odot}$) $\leq$ -1.1). \citet{2001AJ....122..432R} find that the bimodal IMF may be attributed to the expulsion of low-mass objects from small clusters of protostars or from the fragmentation of circumstellar discs. A secondary peak towards the brown dwarf regime may suggest different formation mechanisms for stars and brown dwarfs. We discuss further the dependence of brown dwarf population on stellar density and UV flux across various star forming regions. %We conclude that sub-stellar formation is analogous to star formation mechanism and stellar mass distribution in a cluster is independent of the cluster environment. This result thus, implies a common formation mechanism for both stellar and sub-stellar sources. The top-heavy IMF obtained in this study also indicates a lack of low mass ($<$ 0.2 M$_\odot$) and sub-stellar sources compared to the stars $\geq$ 0.2 M$_\odot$ in IC 1396. The earlier X-ray studies of the target region by \citet{2009AJ....138....7M, 2012MNRAS.426.2917G, 2021AJ....162..279S} using Chandra and XMM data, too support this scenario as the observed X-ray luminosity function for IC 1396 declines rapidly for log L$_{X} >$ 30.5 ergs/s, as compared to other stronger feedback affected regions like ONC (\citealt{2005ApJS..160..319G}), NGC 2244 (\citealt{2002A&A...384..890B}), NGC 3603 (\citealt{2002ApJ...573..191M}). However, whether the amount of X-ray emission observed implies anything definite about the brown dwarf fraction in a region is not clear as of yet.\\ %Since, pre-main sequence stars are strong emitters of X-rays, the less amount of X-ray luminosity detected in the region clearly indicates a dearth of sub-stellar sources (\citealt{2007MNRAS.377.1647N}). 
%This result is well supported by the high star-to-brown dwarf ratio ($\sim$ 5) derived for IC 1396 compared to the strong erfeedback impacted regions like ONC, NGC 2244 and NGC 3603.\\

% This may be the consequence of UV flux generated by the central O-type star which, although impacts the cluster environment but is insufficient to increase the brown dwarf formation by photo-ionisation. In addition to this, a moderate stellar density (25 stars/pc$^{2}$) in the region is inadequate to enhance the brown dwarf population by turbulent fragmentation or disk fragmentation process as compared to the greater brown dwarf census (R $\sim$ 2--3) observed in regions with stronger stellar feedback like ONC, Arches cluster, NGC 2244 ((\citealt{2011A&A...534A..10A, 2020ApJ...896...80G})). 

\subsection{Comparison with other star forming regions}
\label{sec:compare}

\begin{table*}
	\centering
	\begin{tabular}{ l l l l l } % four columns, alignment for each
		\hline
		S.no. & Cluster name & Star-to-brown dwarf-ratio & Surface density & log (flux$_{FUV}$) \\
              &              &                           & (parsec$^{-2}$)     &     (G$_{0}$) \\
		\hline
        \\
		1 & $\sigma$ Ori & 5.4 $\pm$ 1.0 & 36 & 4.00 \\ 
        \\  
        2 & IC 348 & 3.4 $\pm$ 0.6 & 200 & 2.61 \\
        \\
        3 & NGC 1333 & 2.1 $\pm$ 0.2 & 185 & 3.00 \\
        \\
        4 & Chameleon I & 4.0 $\pm$ 0.5 & 30 & 0.90 \\
        \\
        5 & Lupus 3 & 4.3 $\pm$ 0.5 & 20 & 0.70 \\
        \\
        6 & ONC & 2.4 $\pm$ 0.2 & 350 & 5.50 \\
        \\
        7 & NGC 2244 & 2.2 $\pm$ 0.3 & 28 & 5.74 \\
        \\
        8 & RCW 38 & 2.0 $\pm$ 0.6 & 2500 & 6.30 \\
        \\
        9 & $\rho$ Oph & 5.1 $\pm$ 0.4 & 79 & 2.00 \\
        \\
        %10 & NGC 2264 & 3.7 $\pm$ 0.2 & 2 & 5.14 \\
        10 & 25 Ori & 5.9 $\pm$ 0.8 & 4 & 4.00 \\
        \\
        11 & NGC 6611 & 4.4 $\pm$ 0.7 & 138 & 6.03 \\
        \\
        %13 & NGC 3603 & 2.5 $\pm$ 0.3 & 4500 & 7.00 \\
        %14 & W3 & 4.0 $\pm$ 0.1 & $\sim$ 315 & 6.00 \\
        12 & Westerlund 1 & 1.5 $\pm$ 0.2 & 295 & 7.32 \\
        \\
        13 & NGC 2024 & 3.5 $\pm$ 0.4 & 80 & 4.80 \\
        \\
        %17 & Quintuplet & 1.5 $\pm$ 0.2 & 880 & 6.50 \\
        %18 & Arches & 1.2 $\pm$ 0.1 & 1464 & 7.20 \\
        14 & Taurus & 4.0 $\pm$ 0.6 & 5 & 1.30 \\
        \\
        15 & IC 1396 & 6.0 $\pm$ 0.8 & 15 & 5.20 \\
		\hline
	\end{tabular}
    \caption{ shows various star forming regions with their star-to brown dwarf ratio for mass range 0.03 - 1 M$_{\odot}$, stellar surface density and dominant FUV flux. the reference studies for each region according to their serial number is as follows: (1) \citet{2011ApJ...743...64B, 2012ApJ...754...30P, 2022EPJP..137.1132W, 2023ApJ...951..139D} and references therein; (2) \citet{2013ApJ...775..138S, 2019ApJ...881...79M, 2020AJ....160...57L, 2021MNRAS.504.2557D} and references therein; (3) \citet{2013ApJ...775..138S, 2019ApJ...881...79M, 2022MNRAS.510.4888K}; (4),(5) \citet{Mu_i__2015, 2019ApJ...881...79M, 2019PhDT........93W}; (6) \citet{2011A&A...534A..10A, 2019ApJ...881...79M, 2022csss.confE.196K}; (7) \citet{2023A&A...677A..26A, 2019ApJ...881...79M} and references therein ,\citet{2021MNRAS.504.2557D}; (8) \citet{2011ApJ...743..166W, 2017MNRAS.471.3699M, 2019ApJ...881...79M}; (9) \citet{2006ApJ...640..383O, 2012ApJ...744..134M, 2022A&A...667A.163M};% (10) \citet{2008hsf1.book..966D, 2008AJ....135..441S, 2010AJ....140.2070S, 2018A&A...609A..10V, 2021MNRAS.507.4074P, 2023A&A...670A..37F}; 
    (10) \citet{2007ApJ...661.1119B, 2014MNRAS.444.1793D, 2019MNRAS.486.1718S}; (11) \citet{2009MNRAS.392.1034O, 2021MNRAS.504.2557D}; %(13) \citet{1998ApJ...498..278E, 2008ApJ...675.1319H, 2022EPJP..137.1132W} and references therein; (14) \citet{2005AJ....129..393O, 2008ApJ...673..354F, 2009ApJ...693..634O, 2014A&A...561A..12B, 2015ApJ...813...42K, 2021AJ....161..138H};
    (12) \citet{2013AJ....145...46L, 2017A&A...602A..22A, 2022EPJP..137.1132W}; (13) \citet{2006ApJ...646.1215L, 2009MNRAS.392.1034O, 2022EPJP..137.1132W}; %(17) \citet{2000ApJ...545..301K, 2014ApJ...789..115S, 2016MNRAS.460.1854S, 2022EPJP..137.1132W}; (18) \citet{2013ApJ...764..155L, 2015MNRAS.447..366S, 2019ApJ...870...44H, 2022ApJ...939...68H, 2022EPJP..137.1132W}; 
    (14) \citet{2009MNRAS.392.1034O, 2019AJ....158...54E} (15) \citet{2005AJ....130..188S, 2023ApJ...948....7D}} 
    \label{tab:sbd_den_uv}
\end{table*}

We have compiled a list of 15 star forming regions with the related information about their ratio R for mass range 0.03 - 1 M$_{\odot}$, stellar surface density and incident FUV flux in Table \ref{tab:sbd_den_uv}. As evident, the obtained ratio R for IC 1396 is compatible with that estimated for other star forming regions by different studies. The obtained R value is however, at the higher end of the standard range (2 - 6) for the star-to-brown dwarf ratio. As discussed in Section \ref{sec:sbd ratio} this is due to the steep values of R ($>$ 8) observed towards the inner regions ($<$ 3 parsecs) of the cluster. The steep R values are attributed to the higher stellar densities observed in the vicinity of central O-stars as compared to the outskirts, which is a result of sequential star formation episodes (\citealt{2007ApJ...654..316G, 2012AJ....143...61N, 2023ApJ...948....7D}). Also the FUV flux from O-stars inhibits the brown dwarf formation near the cluster center. Although the ratio R decreases with increasing radial distance from the center, when averaged over the area of study it tends towards a higher value. The star-to-brown dwarf ratio is clearly impacted by the presence of OB-stars ratio across the cluster.\\ 

The compiled stellar clusters in Table \ref{tab:sbd_den_uv} are well-studied massive Galactic star forming regions with extreme to low feedback driven cluster environments. These clusters host a stellar density ranging between $\sim$ 4 - 2500 stars pc$^{-2}$ and log (flux$_{FUV}$) $\sim$ 0.7 - 7.3 G$_{0}$. We take the mean value of star-to-brown dwarf ratios for a cluster if a range of values is calculated in the corresponding reference studies which are listed in the caption of Table \ref{tab:sbd_den_uv}. In case of different values of the ratio given in multiple reference studies for any region, we adopt the value determined by the most recent available reference study for that cluster. The error bars represent the error in the ratio for each region, either taken directly from the respective reference study or derived by calculating the Poisson error for the ratio. Also, for some regions like NGC 2024 and Westerlund 1 clusters, we have derived the ratio R from their respective mass functions given in the corresponding reference studies since a direct value was not available (see Table \ref{tab:sbd_den_uv}). For this purpose, we simply calculate number of sources corresponding to the mass function fit performed by the respective authors in the desired mass range (0.03 - 1 M$_{\odot}$). The stellar densities for individual regions were taken directly from the respective reference studies for each region mentioned in the caption of Table \ref{tab:sbd_den_uv}. The regions for which kernel density estimation plots are available in \citet{2019ApJ...881...79M} (e.g. ONC, IC 348, Lupus 3, NGC 1333, RCW 38, NGC 2244 and Cha I), we have considered the stellar density associated with 50$\%$ contour. This is consistent with the methods applied in \citet{2019ApJ...881...79M} to compare the R values among different star forming regions. In order to find the incident FUV flux in each region we use the respective reference studies and obtain the O-stars catalog for each region. We then use the results from \citet{2016arXiv160501773G} and \citet{2022EPJP..137.1132W} to determine the FUV luminosity and thereby, Habing flux\footnote{1 G$_{0}$ = 1.6 x 10$^{-3}$ erg/cm$^{2}$/s} (G$_{0}$) corresponding to the spectral type for O-type stars in individual regions. The FUV fluxes given in Table \ref{tab:sbd_den_uv} are the approximate values as we do not add the B-type star fluxes if the region is dominated by O-type stars. 

It is important to note here that the larger value of star-to-BD ratio R in the cluster centre is not an effect of sample bias. This is because IC 1396 is a very sparse cluster and the central crowding does not affect any source detection. However, for crowded star forming regions (some of them included in Table \ref{tab:sbd_den_uv}), it may be more difficult to detect brown dwarfs in the presence of a concentration of much brighter stars, especially towards the central regions of such clusters. This may have biased the ratio R in such regions towards a higher value.
%For example, \citet{2021MNRAS.507.4074P} find the ratio R $\sim$ 2 - 6 for NGC 2264, the northern sub-cluster of which is powered by a massive O-type star (\citealt{2018A&A...609A..10V}). For example, the ratio R for several nearby star forming regions with varying star forming conditions like $\sigma$ Ori (\citealt{2009A&A...505.1115L, 2012ApJ...754...30P, 2023ApJ...951..139D}), Cha I, Lupus 3 (\citealt{Mu_i__2015}) have been found to vary within 2 - 6. 
%On the contrary, the recent study of NGC 2244 by \citet{2023A&A...677A..26A} observes the brown dwarfs to be more concentrated around the massive stars in the cluster. %of NGC 2244 as $\sim$ 2.2 - 2.8 and find it to be in accordance with six other nearby star forming regions used for comparison by the authors.  The compatibility of the ratio R among Galactic star forming regions implies a similar mass function in the sub-stellar regime across the Milky Way.

\subsection{Impact of cluster environment}
\label{sec:env_impact}
Brown dwarfs form by the gravitational core-collapse of a molecular core which may get deficient of dust and gas due to its expulsion from the material reservoir (\citealt{2017A&A...608A.107V, 2018MNRAS.478.5460R}). The factors leading to this depletion of material range from photo-erosion due to radiative feedback from nearby massive stars, turbulence to disk instability and subsequent ejection of the low-mass core, thus leaving the formed object in sub-stellar regime (\citealt{2009MNRAS.392..413S, 2014prpl.conf..619C, 2018arXiv181106833W}). However, whether the environmental factors play a significant role in deciding the dominant formation mechanism or the census of brown dwarfs in a region is yet to be confirmed. Here we investigate the behaviour of star-to-brown dwarf ratio with respect to the two main factors, that is, stellar density and incident far-UV (FUV) flux impacting the cluster environment.\\ %. Among various formation mechanisms mentioned in Section \ref{sec: intro}, disk fragmentation and 

\begin{figure*}
	%\centering
	%\includegraphics[width=8.5cm, height=6cm]{i2Y_cmd_Yso.png}
	\includegraphics[scale=0.253]{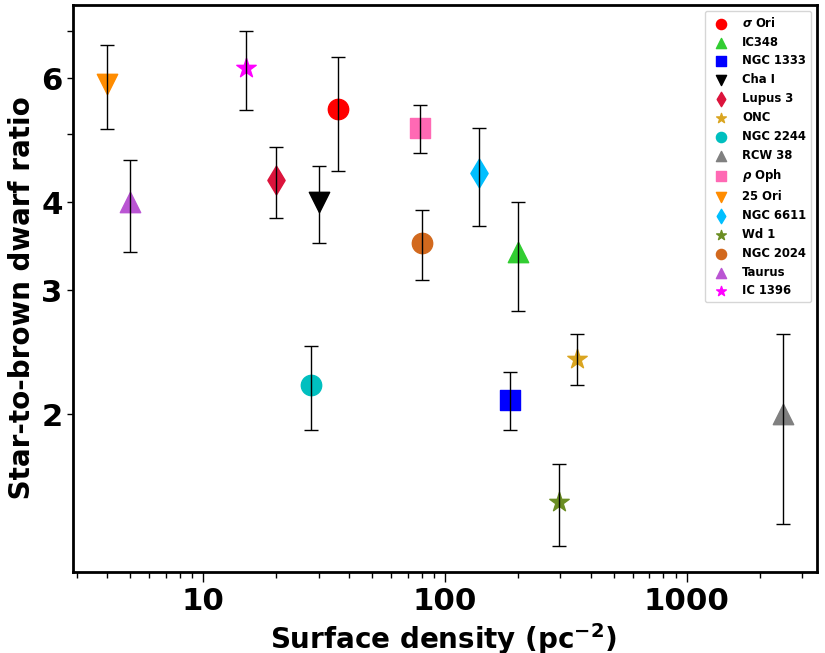}
	\includegraphics[scale=0.252]{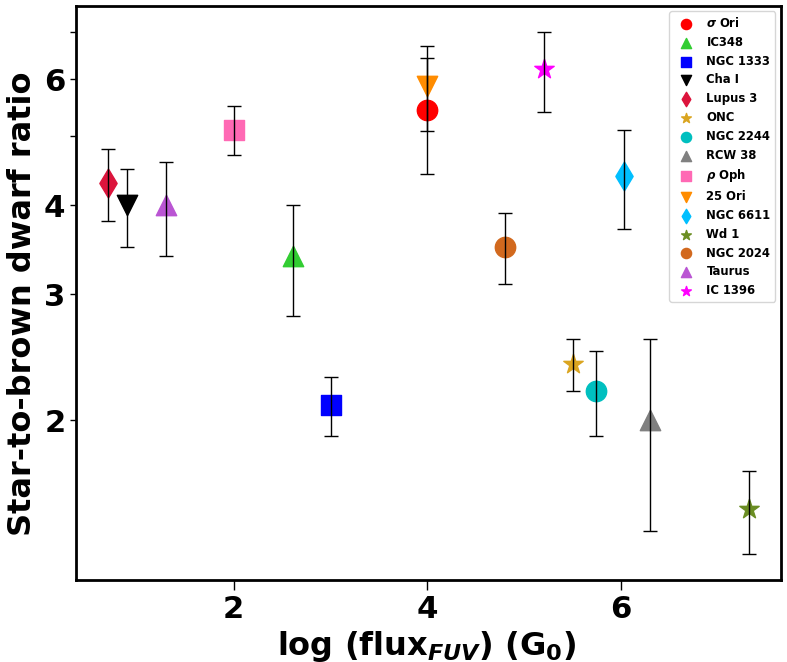}
	%\includegraphics[scale = 0.38]{r2-Y_vs_r2.eps}%\scriptsize
	%\includegraphics[scale = 0.4]{z-K_vs_z.eps}%\scriptsize
	%\begin{small}  
	%\scriptsize
	%\linespread{0.8}
	\caption{ Variation of star-to-brown dwarf ratio versus ({\it Left}) stellar surface density ({\it Right}) incident FUV flux. Different coloured points represent different star forming regions.
	 } 

%{\it Top left:}
% }
%\end{small}
\label{fig: sbd_var}
\end{figure*}

\subsubsection{\it Impact of stellar density} 
\label{sd}
According to theoretical simulations, dynamical interactions among the self-gravitating cores in isolated parts of dense molecular clouds or inside filaments and circumstellar disks may lead to the subsequent ejection of stellar embryos (\citealt{2008MNRAS.389.1556B, 2018arXiv181106833W}). %For example, disk instability leads to fragmentation into several low-mass cores, some of which get ejected out due to the interactions among themselves. 
Such low-mass collapsing cores, depleted of the material reservoir required for accretion eventually form sub-stellar objects (\citealt{2008MNRAS.389.1556B, 2009MNRAS.392..413S, 2014prpl.conf..619C, 2017A&A...608A.107V, 2018arXiv181106833W, 2018MNRAS.478.5460R}). External factors like stellar density may play a crucial role in the pre-mature ejection of sub-stellar cores. For example, close dynamical encounters with nearby stars are a common phenomenon in cluster environments with high stellar density. Such gravitational encounters can enhance the dynamical interactions among the isolated fragments or introduce perturbations in the star-disk system (\citealt{2018MNRAS.478.2700W}) thus, leading to the early ejection of stellar embryos (\citealt{2008MNRAS.389.1556B, 2009MNRAS.392..413S, 2016MNRAS.457..313P, 2017A&A...608A.107V}). \citet{2008MNRAS.389.1556B} find that the fraction of brown dwarfs in a region becomes twice as the stellar density enhances by an order of magnitude. The authors hence hypothesize that high density favours brown dwarf formation. Similarly \citet{2017A&A...608A.107V} conclude that prograde close encounter can result into early ejection of fragments of brown dwarf mass regime. \citet{2018MNRAS.478.5460R} also find a moderate impact of density perturbations on the formation of brown dwarfs and very low mass stars. In the recent studies by \citet{2017MNRAS.471.3699M, 2019ApJ...881...79M}, the authors perform observational analysis of RCW 38 and NGC 2244, the two cluster environments with remarkably contrasting stellar densities, but are inconclusive about any variation in the brown dwarf formation efficiency with respect to either stellar density or presence of OB stars. The most recent study by \citet{2023A&A...677A..26A} also supports this result and does not find any dependence of the star-to-brown dwarf ratio on the stellar density. However, these studies lack a comprehensive analysis for a miscellany of star forming regions and are restricted to inner regions of the respective clusters.\\

We check the variation of star-to-brown dwarf ratio with respect to stellar density across 15 star forming regions in Figure \ref{fig: sbd_var} ({\it Left}). The error bars on each point denote the uncertainty in star-to-brown dwarf ratio for the respective star forming region. %We take the mean value star-to-brown dwarf ratios for the regions if a range of values is calculated in the corresponding reference study. Also, for some regions like NGC 2024, W3, Westerlund 1, NGC 3603, Arches and Quintuplet clusters, we have derived the ratio R by extrapolating the IMF slope given in the corresponding reference studies (see Table \ref{tab:sbd_den_uv}).
We observe that the star-to-brown dwarf ratio tends towards lower values at high stellar densities which, agrees with the theoretical hypothesis that high stellar densities favour the formation of brown dwarfs (\citealt{2008MNRAS.389.1556B}). %This behaviour is however, prominent at very high densities ($>$ 500 stars/pc$^{2}$) and is not distinctly observed at lower densities. 
This behaviour is however, more conspicuous at stellar densities $>$ 50 stars/parsec$^{2}$, which suggests that stellar densities ought to be sufficiently high so as to promote dynamical encounters and effectuate the ejection of sub-stellar cores. We also calculate the correlation coefficient to support our results. Since from Figure \ref{fig: sbd_var} ({\it Left}), the star-to-brown dwarf ratio and stellar density appear to be monotonically related, we evaluate the Spearman's rank correlation coefficient ($\rho$). We find the coefficient $\rho$, to be $\sim - 0.7$ which indicates a strong correlation between the two variables, with a statistical significance of 0.5$\%$ which emphasizes the relation to be monotonic as well (\citealt{10.2307/1422689}). %We find that an increase in the stellar density by a factor of $\sim$ 4 enhances the number of brown dwarfs by a factor of $\sim$ 2. However, this quantification is not definitive as a similar enhancement in the census of brown dwarfs (by a factor of 2--3) is perceived even with a rise in stellar density by an order of magnitude (\citealt{2008MNRAS.389.1556B}). 
We hereby conclude that the brown dwarf formation efficiency may get impacted by high stellar densities ($>$ 50 stars/parsec$^{-2}$) but do not quantify these results as it is beyond the scope of this study.

\subsubsection{\it Impact of FUV flux} 
\label{uv}
Photo-erosion of self-gravitating cores in the vicinity of massive OB-stars is another mechanism of formation of brown dwarfs. The incident FUV and EUV radiation from the surrounding massive stars ionises and erodes the infalling material before it can accrete onto the central core, thus resulting in the formation of a sub-stellar object (\citealt{2004A&A...427..299W}). However, whether this is a dominant formation mechanism or not is still ambiguous. For example, the recent study by \citet{2023A&A...677A..26A} does not find any direct dependence of the brown dwarf formation on OB-stars but, observes brown dwarfs to be located in close proximity to OB stars in NGC 2244. The study is thus, inconclusive regarding the role of OB-stars on brown dwarf formation. In our present study, we try to explore the variation of star-to-brown dwarf ratio relative to the presence of massive stars. %whereas on the contrary, \citet{2017MNRAS.471.3699M, 2019ApJ...881...79M} does not find any effect of the presence of OB stars on the brown dwarf fraction in a region. 
We use the compiled list in Table \ref{tab:sbd_den_uv} to plot the variation of star-to-brown dwarf ratio R with the incident FUV flux in Figure \ref{fig: sbd_var} ({\it Right}), along with the respective error bars representing error in star-to-brown dwarf ratio for various star forming regions.\\

Looking at Figure \ref{fig: sbd_var} ({\it Right}), we are unable to make out a distinct correlation between the incident flux and star-to-brown dwarf ratio. The ratio however, appears to decline with increasing flux at log(flux$_{FUV}$) $>$ 4 G$_{0}$. This could likely be the effect of photo-erosion at high UV-fluxes or high stellar density observed in these regions since both the factors, that is high stellar density and high incident UV flux cannot be decoupled in massive star clusters. We are hence inconclusive regarding the effect of the presence of OB stars on brown dwarf formation efficiency. This result is also supported by \citet{2019ApJ...881...79M} and \citet{2023A&A...677A..26A}, who do not find any clear dependence of brown dwarf fraction on the presence of OB stars based on the comparison among seven star forming regions. This suggests that photo-erosion may not be the dominant mode of brown dwarf formation. However, more comprehensive sub-stellar studies of massive clusters with rich OB-stars population like Cygnus OB2, NGC 3603, W3/W4/W5, Arches and Quintuplet cluster etc. are required to draw any conclusive results.

We caution the readers that although for the same mass range (0.03 - 1 M$_{\odot}$), our comparative analysis however covers different areas of study considered in the respective reference studies to calculate the ratio R in different star forming regions. Another obvious caveat in such comparisons with literature values is the method by which the masses are estimated in each study. We do not consider the mass segregation effect or variation in spatial distribution of brown dwarfs as it is beyond the scope of this study.\\%We do emphasize from our study that FUV flux needs to be sufficiently high to support the formation of brown dwarfs in a cluster. It implies that brown dwarf formation efficiency does increase at high incident FUV fluxes which, might suggest photo-erosion as a crucial formation mechanism in strong feedback impacted cluster environments which confers with the theoretical studies like \citet{2004A&A...427..299W, 2005AN....326..899W}. However, we are unable to draw any conclusive quantitative results at this stage regarding the effect of presence of OB stars on the brown dwarf fraction. 
 %does not effect the  which hence implies that photo-erosion may not be the dominant brown dwarf formation mechanism in a region.
\section{Summary and Future Works}
\label{sec: sumup}
Brown dwarfs form the bridge between stars and planets and hence, it is crucial to study the nature of mass distribution down to sub-stellar regime. The effect of cluster environment on the brown dwarf formation efficiency is another crucial aspect which still requires comprehensive studies, both theoretical and observational. We study the central 22$^{\prime}$ radius region of IC 1396, a nearby ($<$ 1 kpc) prominent \hii region with feedback driven cluster environment ideal for our study.\\

1) We observed 2 pointings each of 1.5$^{\circ}$ diameter centred at IC 1396 with Subaru HSC in r$_{2}$, i$_{2}$ and Y filters. We however perform a deep analysis of the central 22$^{\prime}$ radius region as it covers $\sim$ 60$\%$ of the previously identified sources. This is the deepest study of this region as of yet with detected sources reaching down to r$_{2}$ $\sim$ 28 mag.\\

2) We perform a multi-wavelength study of the region using Subaru HSC, Gaia, Pan-STARRS and UKIDSS / 2MASS data and obtain the candidate members using machine learning tools such as GMM and RF classifier. We then perform SED fitting of the probable members using VOSA tool. We also define the pre-main sequence locus and include sources within 1$\sigma$ as members. Eventually we obtain 458 good quality sources along with their stellar parameters as cluster members.\\

3) Among these members we identify 62 sources as brown dwarfs with mass $\leq$ 0.08 M$_{\odot}$ and obtain the mass distribution down to sub-stellar regime for a mass range 0.03 - 1 M$_{\odot}$. The star-to-brown dwarf ratio R turns out to be $\sim$ 6.0 $\pm$ 0.8 which is compatible with the range found for other star forming regions for the same mass range.\\

4) We find that within IC 1396, the brown dwarf fraction increases with increasing radial distance from the central massive star HD 206267. The O-stars present in IC 1396 influence the star-to-brown dwarf ratio across the region.\\

5) We perform a comprehensive analysis of 15 star forming regions to test the influence of environmental factors like stellar density ($\sim$ 4-2500 stars pc$^{-2}$) and incident FUV flux ($\sim$ 0.7-7.3 G$_{0}$), on the star-to-brown dwarf ratio. We conclude that brown dwarf formation efficiency may be affected by high stellar densities but the effect of the incident FUV flux within this range is unclear.\\

As the next step, we aim to perform a similar analysis for the entire observed region of IC 1396 and obtain the sub-stellar mass distribution for the complex. We also plan to identify the young stellar objects in the region and perform a circumstellar disk analysis to understand the variation of disk fraction with external factors like stellar density and incident FUV flux across the cluster.\\

\section{Data Availability}
\label{sec: bd_table}
Table 3 presents the optical HSC photometry along with the mass of the brown dwarfs identified in this study. The complete catalog of the identified members underlying this article will be shared on reasonable request to the corresponding author.\\

\onecolumn
\begin{longtable}[c]{|ccccccccc|} % four columns, alignment for each
    \caption{Brown dwarfs within the central 22$^{\prime}$ radius region identified in this study. The columns present the RA, Dec, photometry and the respective errors in r$_{2}$, i$_{2}$ and Y filters and the mass of these sources.}\\
	\hline
	RA & Dec & r$_{2}$ & r$_{2_{err}}$ & i$_{2}$ & i$_{2_{err}}$ & Y & Y$_{err}$ & Mass \\
    (deg) & (deg) & (mag) & (mag) & (mag) & (mag) & (mag) & (mag) & (M$_{\odot}$) \\
	\hline
    \endfirsthead

    \hline
	RA & Dec & r$_{2}$ & r$_{2_{err}}$ & i$_{2}$ & i$_{2_{err}}$ & Y & Y$_{err}$ & Mass \\
    (deg) & (deg) & (mag) & (mag) & (mag) & (mag) & (mag) & (mag) & (M$_{\odot}$) \\
	\hline
    \endhead
    \hline
        325.0996 & 57.7831 & 23.74 & 0.01 & 20.61 & 0.00 & 18.34 & 0.00 & 0.07\\
        325.2094 & 57.7277 & 24.22 & 0.01 & 20.30 & 0.00 & 18.57 & 0.00 & 0.06\\
        324.7828 & 57.4264 & 24.07 & 0.01 & 20.56 & 0.00 & 18.91 & 0.00 & 0.06\\
        324.0923 & 57.4959 & 23.85 & 0.01 & 20.54 & 0.00 & 18.78 & 0.00 & 0.06\\
        324.4811 & 57.4484 & 23.69 & 0.01 & 20.11 & 0.00 & 18.40 & 0.00 & 0.06\\
        324.4575 & 57.4969 & 26.06 & 0.06 & 21.49 & 0.00 & 19.65 & 0.00 & 0.04\\
        324.3343 & 57.7497 & 25.55 & 0.03 & 21.08 & 0.00 & 19.01 & 0.00 & 0.03\\
        324.1520 & 57.5092 & 24.45 & 0.02 & 20.98 & 0.00 & 19.29 & 0.00 & 0.06\\
        324.1435 & 57.5154 & 24.92 & 0.03 & 21.04 & 0.00 & 19.30 & 0.00 & 0.06\\
        325.1902 & 57.5676 & 24.23 & 0.01 & 20.86 & 0.00 & 19.17 & 0.00 & 0.06\\
        324.4685 & 57.3126 & 22.74 & 0.00 & 20.11 & 0.00 & 18.40 & 0.00 & 0.07\\
        324.3183 & 57.5488 & 22.72 & 0.00 & 19.96 & 0.00 & 18.33 & 0.00 & 0.07\\
        324.1370 & 57.5015 & 23.51 & 0.01 & 20.74 & 0.00 & 19.22 & 0.00 & 0.06\\
        324.2120 & 57.6214 & 22.73 & 0.00 & 20.01 & 0.00 & 18.50 & 0.00 & 0.07\\
        325.3353 & 57.3186 & 22.79 & 0.00 & 20.24 & 0.00 & 18.70 & 0.00 & 0.07\\
        325.3433 & 57.3237 & 22.85 & 0.00 & 20.31 & 0.00 & 18.76 & 0.00 & 0.07\\
        325.1490 & 57.2933 & 22.50 & 0.00 & 20.03 & 0.00 & 18.64 & 0.00 & 0.07\\
        325.1416 & 57.3540 & 23.36 & 0.00 & 20.63 & 0.00 & 19.06 & 0.00 & 0.06\\
        324.9703 & 57.4901 & 22.66 & 0.00 & 19.86 & 0.00 & 18.29 & 0.00 & 0.07\\
        325.0021 & 57.5585 & 22.83 & 0.00 & 20.22 & 0.00 & 18.69 & 0.00 & 0.07\\
        325.1468 & 57.6155 & 22.80 & 0.00 & 20.27 & 0.00 & 18.65 & 0.00 & 0.07\\
        324.9589 & 57.3731 & 23.33 & 0.01 & 20.57 & 0.00 & 18.78 & 0.00 & 0.07\\
        324.9279 & 57.4383 & 22.71 & 0.00 & 20.09 & 0.00 & 18.38 & 0.00 & 0.07\\
        324.9290 & 57.4600 & 23.03 & 0.00 & 20.42 & 0.00 & 18.69 & 0.00 & 0.07\\
        324.9169 & 57.5325 & 23.47 & 0.01 & 20.75 & 0.00 & 18.98 & 0.00 & 0.06\\
        324.5742 & 57.2674 & 23.91 & 0.01 & 21.04 & 0.00 & 19.34 & 0.00 & 0.06\\
        324.3852 & 57.4135 & 24.95 & 0.02 & 21.79 & 0.00 & 19.66 & 0.00 & 0.06\\
        324.4519 & 57.4703 & 23.22 & 0.01 & 20.14 & 0.00 & 18.74 & 0.00 & 0.07\\
        324.4618 & 57.4889 & 23.83 & 0.01 & 20.57 & 0.00 & 18.60 & 0.00 & 0.06\\
        324.5109 & 57.4804 & 24.43 & 0.02 & 20.87 & 0.00 & 19.16 & 0.00 & 0.06\\
        324.5797 & 57.5147 & 23.87 & 0.01 & 20.88 & 0.00 & 19.30 & 0.00 & 0.06\\
        324.4841 & 57.5443 & 23.76 & 0.01 & 20.54 & 0.00 & 18.70 & 0.00 & 0.06\\
        324.1735 & 57.5382 & 23.53 & 0.01 & 20.54 & 0.00 & 18.32 & 0.00 & 0.06\\
        324.2134 & 57.5346 & 24.34 & 0.01 & 20.91 & 0.00 & 19.15 & 0.00 & 0.06\\
        324.1619 & 57.5102 & 25.72 & 0.06 & 21.93 & 0.00 & 19.89 & 0.00 & 0.06\\
        324.1149 & 57.6079 & 24.68 & 0.02 & 21.06 & 0.00 & 19.43 & 0.00 & 0.06\\
        324.9834 & 57.2734 & 23.68 & 0.01 & 20.78 & 0.00 & 18.97 & 0.00 & 0.06\\
        325.2606 & 57.3339 & 23.23 & 0.01 & 20.25 & 0.00 & 18.51 & 0.00 & 0.07\\
        325.1878 & 57.4292 & 24.15 & 0.01 & 21.15 & 0.00 & 19.25 & 0.00 & 0.06\\
        325.0855 & 57.3392 & 24.07 & 0.01 & 21.01 & 0.00 & 19.30 & 0.00 & 0.06\\
        325.0216 & 57.5352 & 24.29 & 0.01 & 20.83 & 0.00 & 19.29 & 0.00 & 0.06\\
        325.0511 & 57.6162 & 23.92 & 0.01 & 20.96 & 0.00 & 19.13 & 0.00 & 0.06\\
        325.1674 & 57.6198 & 24.99 & 0.02 & 21.74 & 0.00 & 19.95 & 0.00 & 0.06\\
        325.1083 & 57.7926 & 23.97 & 0.01 & 21.03 & 0.00 & 18.95 & 0.00 & 0.06\\
        325.0825 & 57.7862 & 26.36 & 0.06 & 22.62 & 0.00 & 20.49 & 0.00 & 0.05\\
        324.7593 & 57.1508 & 23.92 & 0.01 & 20.67 & 0.01 & 18.82 & 0.00 & 0.06\\
        324.8858 & 57.2525 & 22.92 & 0.00 & 20.08 & 0.00 & 18.42 & 0.00 & 0.07\\
        324.9626 & 57.4817 & 23.20 & 0.01 & 20.32 & 0.00 & 18.49 & 0.00 & 0.07\\
        324.9151 & 57.5068 & 24.70 & 0.02 & 21.11 & 0.00 & 18.81 & 0.00 & 0.06\\
        324.8126 & 57.5849 & 23.38 & 0.01 & 20.50 & 0.00 & 18.70 & 0.00 & 0.06\\
        324.9211 & 57.6997 & 25.34 & 0.03 & 21.72 & 0.00 & 19.79 & 0.00 & 0.06\\
        324.8863 & 57.7422 & 24.23 & 0.01 & 20.83 & 0.00 & 19.24 & 0.00 & 0.06\\
        324.8643 & 57.2352 & 22.46 & 0.00 & 19.77 & 0.00 & 18.38 & 0.00 & 0.07\\
        324.5316 & 57.5287 & 22.50 & 0.00 & 19.37 & 0.00 & 17.60 & 0.00 & 0.07\\
        324.5085 & 57.6752 & 23.20 & 0.01 & 19.79 & 0.00 & 17.90 & 0.00 & 0.07\\
        324.2213 & 57.5187 & 23.68 & 0.01 & 20.08 & 0.00 & 18.42 & 0.00 & 0.06\\
        324.2461 & 57.4858 & 22.46 & 0.00 & 18.80 & 0.00 & 17.05 & 0.00 & 0.07\\
        324.1375 & 57.4803 & 22.25 & 0.00 & 18.85 & 0.00 & 17.32 & 0.00 & 0.08\\
        324.7057 & 57.4316 & 22.33 & 0.00 & 19.41 & 0.00 & 18.10 & 0.00 & 0.08\\
        324.6640 & 57.6054 & 22.58 & 0.00 & 19.61 & 0.00 & 18.23 & 0.00 & 0.07\\
        325.1605 & 57.4679 & 23.16 & 0.00 & 18.91 & 0.00 & 17.42 & 0.00 & 0.07\\
        324.6305 & 57.4982 & 22.31 & 0.00 & 19.68 & 0.00 & 18.08 & 0.00 & 0.08\\
    \hline    
%\label{tab:data}
\end{longtable}
\label{tab: bd_data}
\twocolumn
%%%%%%%%%%%%%%%%%%%%%%%%%%%%%%%%%%%%%%%%%%

\section*{Acknowledgements}
The authors thank the referee for the useful constructive comments which has refined the overall structure, quality and comprehensibility of this paper. This research is based on data collected at Subaru Telescope with Hyper Suprime-Cam, which is operated by the National Astronomical Observatory of Japan. We are honored and grateful for the opportunity of observing the Universe from Mauna Kea, which has the cultural, historical and natural significance in Hawaii. We are grateful to The East Asian Observatory which is supported by The National Astronomical Observatory of Japan; Academia Sinica Institute of Astronomy and Astrophysics; the Korea Astronomy and Space Science Institute; the Operation, Maintenance and Upgrading Fund for Astronomical Telescopes and Facility Instruments, budgeted from the Ministry of Finance (MOF) of China and administrated by the Chinese Academy of Sciences (CAS), as well as the National Key R\&D Program of China (No. 2017YFA0402700). We use Pan-STARRS and GAIA DR3 data for the membership analysis in this work. The Pan-STARRS1 Surveys (PS1) and the PS1 public science archive have been made possible through contributions by the Institute for Astronomy, the University of Hawaii, the Pan-STARRS Project Office, the Max-Planck Society and its participating institutes, the Max Planck Institute for Astronomy, Heidelberg and the Max Planck Institute for Extraterrestrial Physics, Garching, The Johns Hopkins University, Durham University, the University of Edinburgh, the Queen's University Belfast, the Harvard-Smithsonian Center for Astrophysics, the Las Cumbres Observatory Global Telescope Network Incorporated, the National Central University of Taiwan, the Space Telescope Science Institute, the National Aeronautics and Space Administration under Grant No. NNX08AR22G issued through the Planetary Science Division of the NASA Science Mission Directorate, the National Science Foundation Grant No. AST-1238877, the University of Maryland, Eotvos Lorand University (ELTE), the Los Alamos National Laboratory, and the Gordon and Betty Moore Foundation. This work has made use of data from the European Space Agency (ESA) mission GAIA processed by Gaia Data processing and Analysis Consortium (DPAC: https://www.cosmos.esa.int/web/gaia/dpac/consortium). This publication makes use of VOSA, developed under the Spanish Virtual Observatory (https://svo.cab.inta-csic.es) project funded by MCIN/AEI/10.13039/501100011033/ through grant PID2020-112949GB-I00.
VOSA has been partially updated by using funding from the European Union's Horizon 2020 Research and Innovation Programme, under Grant Agreement nº 776403 (EXOPLANETS-A). We gratefully acknowledge the use of high performance computing facilities at IUCAA, Pune for the HSC data reduction.

The authors thank Kora Mu$\check{z}$i$\Acute{c}$ for the valuable review and suggestions which have improved the quality of the paper overall, Surhud More for his help during HSC data reduction. We thank Gregory J. Herczeg for his valuable help with the observing proposal for Subaru HSC for IC 1396 observations. We thank Isabelle Baraffe for providing us with isochrone models for an interval of log (Age) = 0.01, through personal communication. JJ acknowledges the financial support received through the DST-SERB grant SPG/2021/003850. S.R.D. acknowledges support from FONDECYT Postdoctoral fellowship (project code: 3220162). ZG is supported by the ANID FONDECYT Postdoctoral program No. 3220029. ZG acknowledge support by ANID, -- Millennium Science Initiative Program -- NCN19\_171.

%%%%%%%%%%%%%%%%%%%%%%%%%%%%%%%%%%%%%%%%%%%%%%%%%%



\bibliographystyle{mnras}
\bibliography{example} % if your bibtex file is called example.bib

%\begin{thebibliography}{99}
%\end{thebibliography}

%%%%%%%%%%%%%%%%%%%%%%%%%%%%%%%%%%%%%%%%%%%%%%%%%%

%%%%%%%%%%%%%%%%% APPENDICES %%%%%%%%%%%%%%%%%%%%%

\appendix

\section{Data quality comparison}
\label{sec:gaia_quality}
\begin{figure}
	%\centering
	%\includegraphics[width=8.5cm, height=6cm]{i2Y_cmd_Yso.png}
	%\includegraphics[scale=0.155]{mag_lmt.png}
    \includegraphics[scale=0.22]{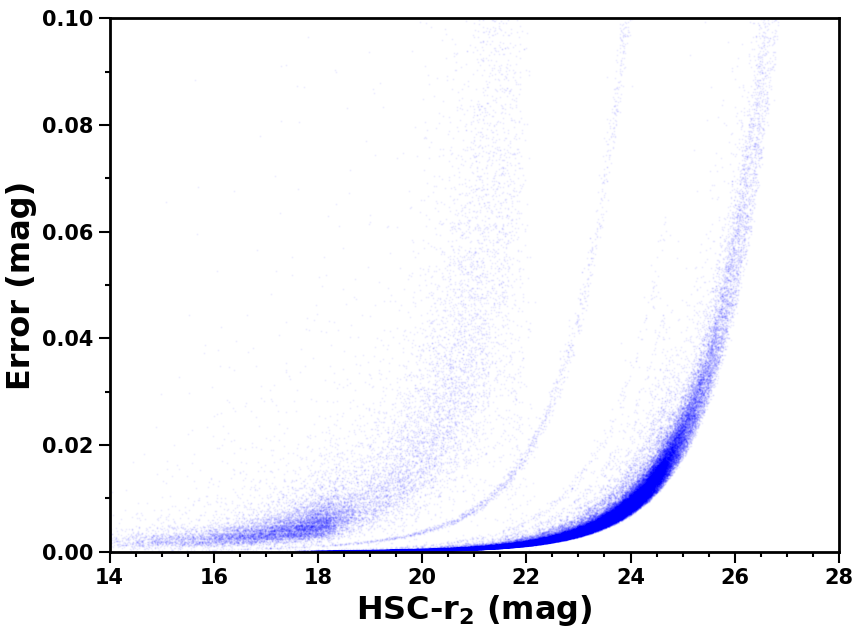}
	%\includegraphics[scale=0.28]{dist_Lit_new.png}
	%\includegraphics[scale = 0.38]{r2-Y_vs_r2.eps}%\scriptsize
	%\includegraphics[scale = 0.4]{z-K_vs_z.eps}%\scriptsize
	%\begin{small}  
	%\scriptsize
	%\linespread{0.8}
	\caption{ ({\it Top}) Completeness of HSC r$_{2}$ filter based on the peak (turnover point) of distribution (\citealt{2017ApJ...836...98J, 2021MNRAS.508.3388G}) ({\it Bottom}) Error vs magnitude for r$_{2}$-filter. The three branches observed in the HSC r$_{2}$-band plot correspond to the merged Pan-STARRS photometry, long and short exposure photometry.
	 } 

%{\it Top left:}
% }
%\end{small}
\label{fig: data_qualty}
\end{figure}

\begin{figure}
	%\centering
	%\includegraphics[width=8.5cm, height=6cm]{i2Y_cmd_Yso.png}
	\includegraphics[scale=0.2]{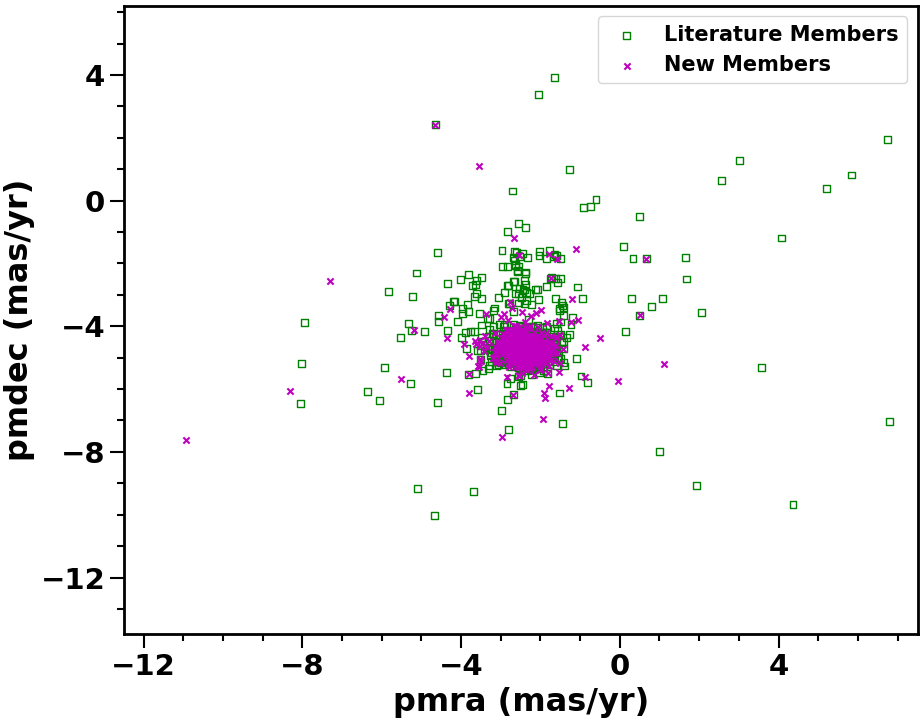}
    \includegraphics[scale=0.23]{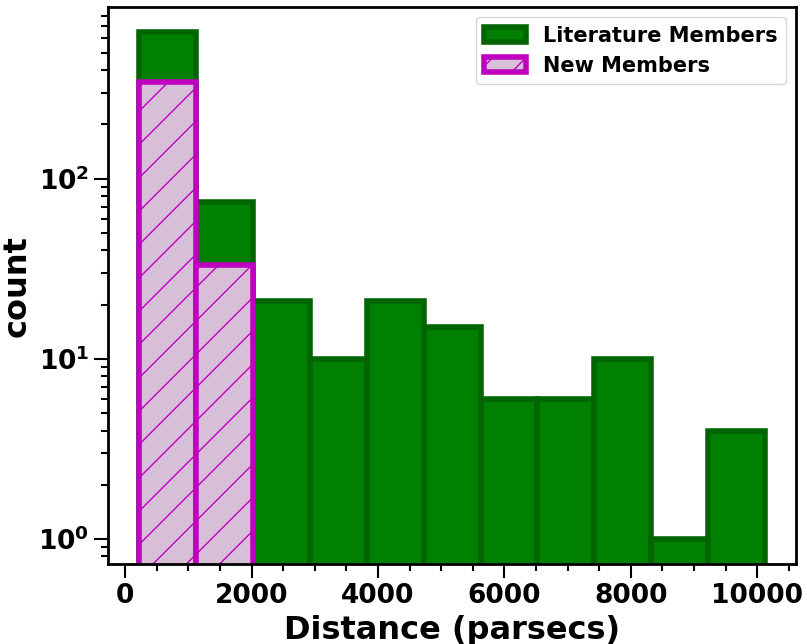}
    %\includegraphics[scale=0.23]{dist_Lit_new.png}
	%\includegraphics[scale=0.28]{dist_Lit_new.png}
	%\includegraphics[scale = 0.38]{r2-Y_vs_r2.eps}%\scriptsize
	%\includegraphics[scale = 0.4]{z-K_vs_z.eps}%\scriptsize
	%\begin{small}  
	%\scriptsize
	%\linespread{0.8}
	\caption{ ({\it Top}) Proper motion RA vs Dec distribution ({\it Bottom}) Distance distribution for cluster members identified by previous studies versus members from the present study with Gaia DR3 counterparts. We observe a small scatter in the proper motions and distance values of the members identified in this study as compared to the literature-based members. However, some scatter is inevitable in all membership analysis. %({\it Right}) Distance distribution for literature based sources and new members with Gaia DR3 counterparts. 
	 } 

%{\it Top left:}
% }
%\end{small}
\label{fig: gaia_qualty}
\end{figure}

In this section, we present two data quality plots of the HSC data used in the present study. As can be seen from Figure \ref{fig: data_qualty} ({\it Top}), the completeness of the data in r$_{2}$-band is down to 24.5 mag ($\sim$ 0.03 M$_{\odot}$). The majority of sources have photometric error $\leq$ 0.05 mag as evident from Figure \ref{fig: data_qualty} ({\it Bottom}). The three branches observed in the HSC r$_{2}$-band plot correspond to the merged Pan-STARRS photometry (used while generating the deep optical catalog as mentioned in Section \ref{sec:ps_data}), long and short exposure photometry. More details about merging of the short and long exposure catalogs will be presented in Das et al., in preparation. We compare the proper motion and parallax distribution of Gaia DR3 counterparts of the members identified in this study with that of the previous studies. Figure \ref{fig: gaia_qualty} {\it Top} and Figure \ref{fig: gaia_qualty} {\it Bottom} suggest an excellent concordance between the proper motion and distance estimates of the new and literature-based cluster members. Also, we observe an unavoidable yet limited scatter in the proper motions and distance values of the members identified in this study as compared to the literature-based members. However, some scatter is inevitable in all membership analysis and has been observed in previous studies (\citealt{2018A&A...618A..93C, 2022arXiv221011930P, 2023ApJ...948....7D}) as well.

\begin{figure}
	%\centering
	%\includegraphics[width=8.5cm, height=6cm]{i2Y_cmd_Yso.png}
	\includegraphics[scale=0.14]{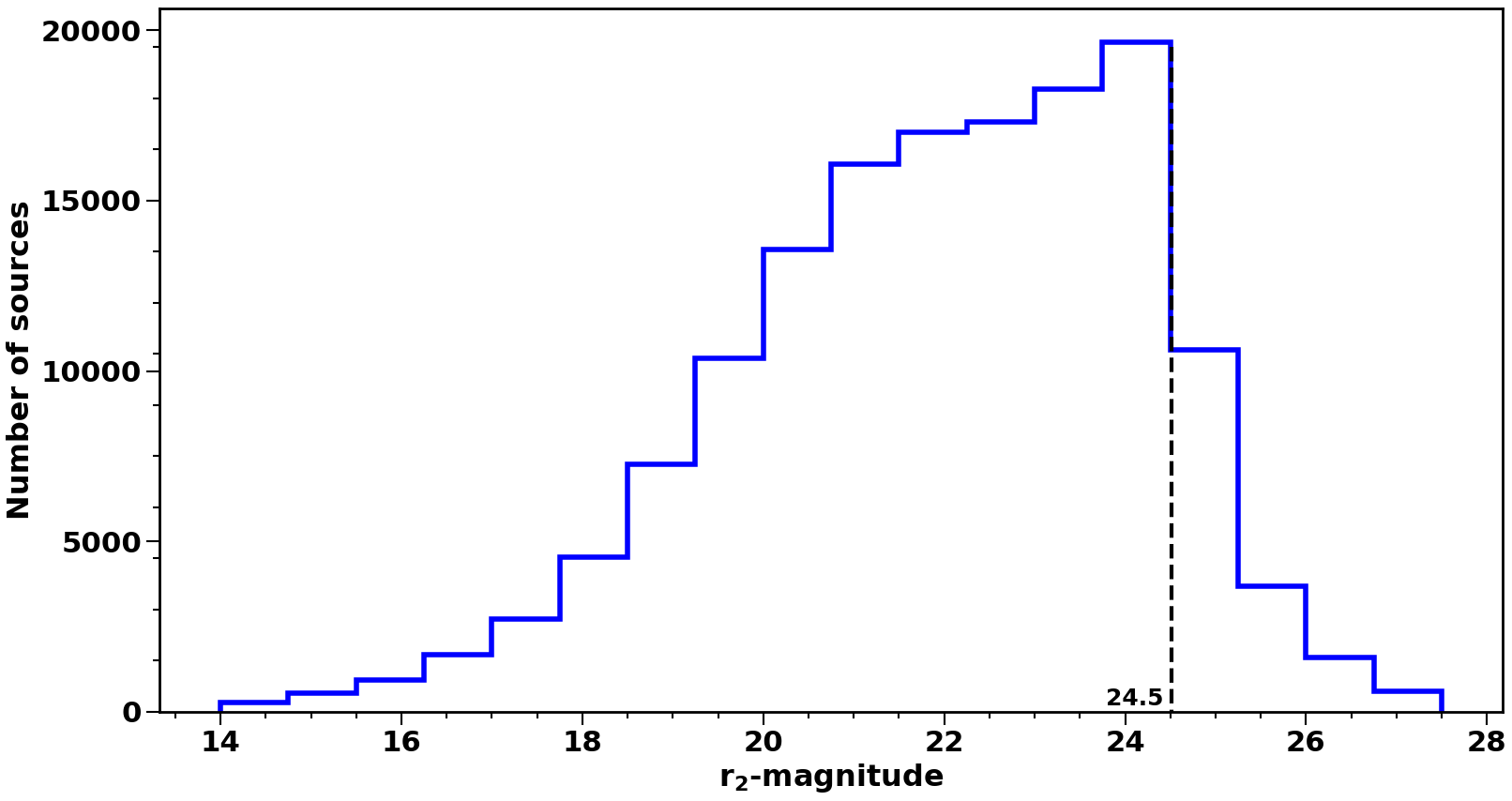}
    %\includegraphics[scale=0.22]{mag_err.png}
	%\includegraphics[scale=0.28]{dist_Lit_new.png}
	%\includegraphics[scale = 0.38]{r2-Y_vs_r2.eps}%\scriptsize
	%\includegraphics[scale = 0.4]{z-K_vs_z.eps}%\scriptsize
	%\begin{small}  
	%\scriptsize
	%\linespread{0.8}
	%\caption{ ({\it Left}) Completeness of HSC r$_{2}$ filter ({\it Right}) error vs magnitude for r$_{2}$-filter. 
	 %} 

%{\it Top left:}
% }
%\end{small}
%\label{fig: data_qualty}
\end{figure}

\begin{figure*}
	%\centering
	%\includegraphics[width=8.5cm, height=6cm]{i2Y_cmd_Yso.png}
	%\includegraphics[scale=0.27]{pm_Lit_new.png}
	\includegraphics[scale=0.22]{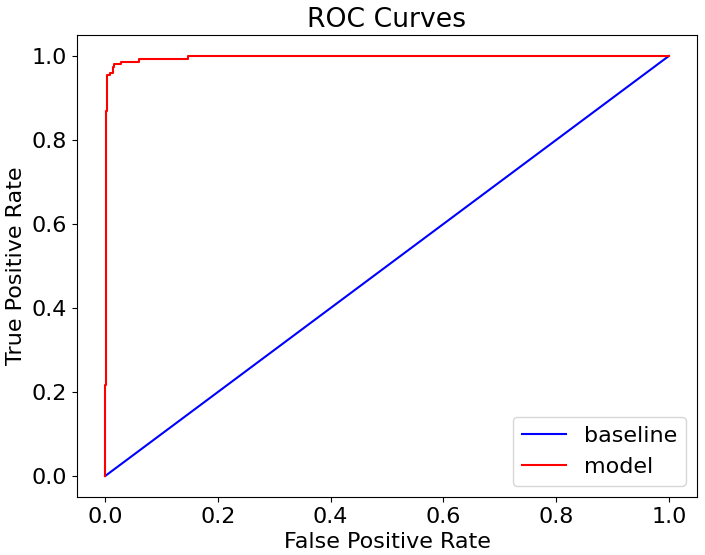}
	\includegraphics[scale = 0.22]{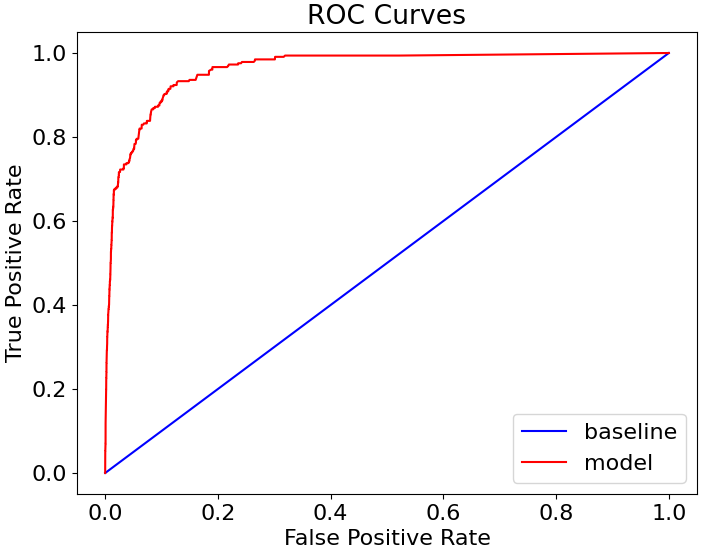}%\scriptsize
	\includegraphics[scale = 0.22]{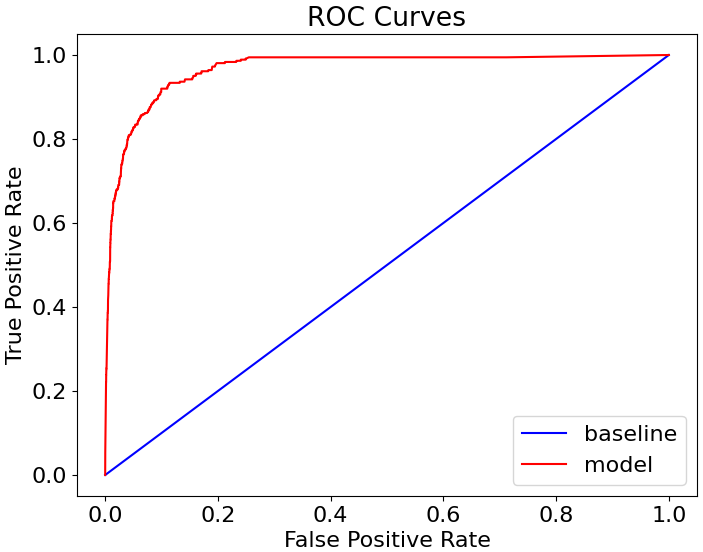}%\scriptsize
	%\begin{small}  
	%\scriptsize
	%\linespread{0.8}
	\caption{ ROC curve for the ({\it Left}) first RF classifier phase. ({\it Center}) second RF classifier phase. ({\it Right}) third RF classifier phase. %Green squares are the literature based sources with Gaia counterparts whereas magenta crosses are Gaia counterparts of members identified in this study.
	 } 

%{\it Top left:}
% }
%\end{small}
\label{fig: rf_qualty}
\end{figure*}

\label{sec:rf_quality}
Hyper-parameters for all the phases of RF classifier are as follows:\\
random state = 50\\ 
number of estimators = 188\\ 
min$\_$samples$\_$split: 5\\ 
max$\_$leaf$\_$nodes: 14\\ 
max$\_$features: 0.6\\ 
max$\_$depth: 16\\ 

We provide all the necessary details of the entire RF process like accuracy, precision, recall, F1 score and CV Score for each phase in Table \ref{tab:data_qual}. We have also included the roc curves for the three phases here.

\begin{table*}
	\centering
	\begin{tabular}{|c|c|c|c|c|c|} %l l l  % four columns, alignment for each
		\hline
		Phase & Accuracy & Precision & F1 score & Recall & CV Score \\
		\hline
		Phase 1 & 0.98 & 0.98 & 0.95 & 0.93 & 0.98 \\ 
        Phase 2 & 0.98 & 0.80 & 0.70 & 0.60 & 0.98 \\
        Phase 3 & 0.98 & 0.72 & 0.70 & 0.62 & 0.98 \\
        \hline
	\end{tabular}
    \caption{ Quality parameters of the three Random Forest classifier phases}
    \label{tab:data_qual}
\end{table*}

\section{Field Subtraction Method for Validation}
\label{sec:fs}

We consider a 10$^{\prime}$ radius control field located towards the outskirts of the HSC observed area of IC 1396 and perform statistical field decontamination of the cluster field, located towards the center of the same area. An area of only 10$^{\prime}$ radius is chosen here for field subtraction due to the lack of a suitable control field of larger area, devoid of young pre-main sequence stars, as observed by their spatial distribution (Das et al., in prep). The 10$^{\prime}$ radius control field considered for the field subtraction here is chosen carefully towards the outskirts as it is devoid of any pre-main sequence sources which is suggested by the spatial density map given in \citealt{2023ApJ...948....7D}. We perform the field decontamination by dividing the r$_{2}$ - Y color and r$_{2}$ magnitude parameter space into 0.1 mag bins, followed by the subtraction of the control field count from the cluster field count (\citealt{2021MNRAS.508.3388G, 2021MNRAS.504.2557D}). The member selection using pre-main sequence locus approach is then performed for the statistically subtracted sources located above the 10 Myr isochrone. Subsequently, sources within 1 $\sigma$ limit of the defined locus and r$_{2}$ $\geq$ 22 mag are selected as members. However, we find that whether or not prior field subtraction is performed, the statistics of members as well as mass distribution remains approximately unaltered. We would like to emphasize here that the sources by field subtraction are not the actual cluster members but just the statistical members in the 10$^{\prime}$ central cluster field. Hence, we continue without field decontamination for pre-main sequence based membership analysis as given in Section \ref{sec:pms locus}. This is because the cluster members obtained within 1 $\sigma$ of the empirical pre-main sequence locus are common in all the CMD combinations (as mentioned in Section \ref{sec:pms locus}). On the contrary, the candidate members selected from the field decontamination by statistical method need not be common across all combinations of different bands as they are not true members. Moreover, due to the lack of a suitable control field of 22$^{\prime}$ radius area, we are unable to perform decontamination for the central cluster considered for our analysis.\\
%%%%%%%%%%%%%%%%%%%%%%%%%%%%%%%%%%%%%%%%%%%%%%%%%%
% Don't change these lines
\bsp	% typesetting comment
\label{lastpage}
\end{document}